# Using backscattered thermal neutrons to monitor boron concentration during BNCT: a Monte Carlo feasibility study

Zirui Ye[a], Yuxin Wang[a], Meitong Wei[a], Xie George Xu[a,*]

[a] School of Nuclear Science and Technology, University of Science and Technology of China, Huangshan Road 443, Hefei, Anhui 230026, China

[*] Corresponding author: Xie George Xu, FAIMBE, FAAPM, FHPS, FANS, University of Science and Technology of China, Huangshan Road 443, Hefei, Anhui, 230026, China.

Email: xgxu@ustc.edu.cn

## Abstract

Boron neutron capture therapy (BNCT) requires knowledge of patient-specific $^{10}$B concentration for accurate dose estimation, yet no established method provides real-time boron-sensitive information during irradiation. Backscattered thermal neutrons carry a $^{10}$B-dependent intensity modulation through the $^{10}$B(n,$\alpha$)$^{7}$Li reaction, documented in BNCT treatment rooms for three decades but not yet developed as a measurement signal. This paper uses Monte Carlo simulation to assess the feasibility of backscattered thermal neutrons as a measurement channel for $^{10}$B concentration. A thin $^{nat}$LiF-converter detector placed at the beam exit captures the composite forward-plus-backscatter field; differential imaging against a $^{10}$B-free baseline isolates the $^{10}$B-dependent component, quantified by the fractional reduction in the $^{6}$Li detector capture rate, termed Relative Detector Signal Reduction (RDSR). In homogeneous phantoms, RDSR shows linear concentration dependence ($R^2 =$ 0.997) with a practical depth limit of approximately 6 cm. Edge-response analysis yields a diffusion-limited FWHM of 32–176 mm over the 1–5 cm depth range, with weak concentration dependence. In a voxelized patient phantom across 12 boron configurations, the $^{6}$Li capture cross-section provides intrinsic thermal neutron energy selectivity that preferentially weights the thermal band where $^{10}$B absorption is concentrated. Region-of-interest integration achieves counting-statistics sensitivity below 10 ppm; the systematic detection floor (~22–28 ppm at ±1% baseline uncertainty) identifies baseline-reference precision as the dominant constraint. Inserting the modeled detector produces limited dose perturbation, with a treatment-time increase of +11.6%. These computational results establish the physical basis for a boron-sensitive backscattered neutron measurement concept in BNCT. Experimental validation and multi-patient generalization remain necessary.

**Keywords:** boron neutron capture therapy, backscattered thermal neutrons, neutron measurement, LiF converter, Monte Carlo simulation, boron concentration

## 1. Introduction

Boron neutron capture therapy (BNCT) delivers a targeted dose through the $^{10}B(n,\alpha)^{7}Li$ reaction, where short-range, high-linear energy transfer (LET) products confine energy deposition to boron-loaded cells Barth et al. (2005). The advent of accelerator-based BNCT (AB-BNCT) has moved the modality from research reactors to hospital-based clinical practice Suzuki (2020), supported by early real-world data Hirose and Sato (2024) and the first randomized controlled trial Kashiwagi et al. (2026). Across all reported series, treatment planning relied on assumed $^{10}B$ concentration ratios, with no real-time boron monitoring available during irradiation. The therapeutic dose is a superposition of boron, nitrogen, fast neutron, and gamma components, each weighted by tissue-specific biological effectiveness factors. The local $^{10}B$ concentration is therefore a primary patient-specific variable in dose estimation. Treatment planning systems (TPS) for BNCT typically assume a fixed tumor-to-blood concentration ratio, but actual tumor uptake varies widely — typically two to four times the blood concentration, with wide inter-patient variability Coderre and Morris (1999). The resulting uncertainty propagates directly into the delivered dose. Spatially resolved measurement of patient-specific $^{10}B$ concentration during treatment delivery remains a major radiation-measurement challenge in BNCT IAEA (2023).

Existing approaches to patient boron assessment each address a distinct aspect of the problem. For examples, $^{18}F$-fluoroboronophenylalanine ($^{18}F$-FBPA) positron emission tomography provides spatial boron distribution but requires a separate pre-treatment scan session with pharmacokinetic extrapolation from microdose tracer data to the therapeutic $^{10}B$ concentration at irradiation time Shimosegawa et al. (2016). Prompt-gamma imaging of the 478 keV $^{10}$B de-excitation line has been demonstrated in principle, but CdTe cameras required 2–20 hours for adequate statistics at a research-reactor facility with a flux one to two orders of magnitude below clinical accelerator levels Chiu et al. (2025). Quantitative 478 keV prompt-gamma imaging has been demonstrated using an electron-tracking Compton camera at a clinical BNCT irradiation facility, achieving spatial imaging of 46–93 ppm boron solutions; however, the system-to-target distance exceeded 2.5 m and the spatial resolution was limited to approximately 10 cm Mizumoto et al. (2025). Offline activation foils provide absolute thermal neutron fluence with 4.5–4.6% expanded uncertainty but require dedicated irradiation followed by offline gamma spectrometry, precluding delivery-time feedback Suzuki et al. (2024). Delivery-time beam monitors — including fission chambers, scintillating optical-fiber detectors, and LiF-based active neutron detectors — measure neutron fluence rate in real time at fixed locations within or near the beam-shaping assembly (BSA) but provide no boron-sensitive spatial information Kumada et al. (2024). Takada et al. developed a thin Si+LiF detector for online neutron beam monitoring at the neutron source target unit, achieving 1.5% fluence-rate precision at $\geq$ 1 s integration; this detector provides a single-point scalar readout without spatial or boron-concentration information Takada et al. (2025). Lithium-based optical imaging provides rapid two-dimensional thermal neutron maps in forward-beam phantom configurations Yamamoto et al. (2022). A $^{10}$B-loaded liquid-scintillator method has demonstrated direct boron-dose profiling along the beam axis, but acquires only discrete depth points through sequential scanning rather than a simultaneous two-dimensional map Maeda et al. (2024). Indirect neutron radiography using activation foils and imaging plates has been validated for two-dimensional thermal neutron reaction-rate mapping in BNCT beam characterization, with results within 5% of independent activation-foil measurements at depths up to 4 cm in the forward-beam direction Wang et al. (2024); however, the method is offline, requires post-irradiation processing, and measures beam-quality parameters rather than patient boron concentration. Transmission-mode neutron radiography has been applied to boron concentration measurement in solution samples, but the authors noted that the technique is not suitable for concentrations below 1000 ppm Dastjerdi et al. (2026); this threshold is well above the concentration scale relevant to BNCT boron monitoring. None of the above methods combines boron-sensitive spatial mapping with real-time operation at clinical concentration levels. This gap motivates investigation of an alternative signal carrier — patient-emergent backscattered thermal neutrons — that may combine spatial mapping capability with real-time operation using a thin, neutron-transparent detector, as addressed in the present study.

A complementary information channel has been documented in BNCT treatment rooms for nearly three decades, yet has been treated as a nuisance, correction term, or confounding factor rather than a usable signal. In 1996, Raaijmakers et al. observed that backscattered thermal neutrons — primarily from the collimator aperture assembly and, to a lesser extent, from the phantom — perturbed monitor count rates, requiring careful calibration under each beam configuration Raaijmakers et al. (1996). Liu et al. separately installed fission chambers inside the collimator explicitly to avoid backscattering contributions Liu et al. (2011). Most recently, Jakubowski et al. reported that the backscatter effect varies with $^{10}$B concentration, but treated this dependence as a confounding factor requiring detector-specific correction, recommending against patient-surface deployment Jakubowski et al. (2024). Prior BNCT thermal neutron imaging concepts such as the Thermal Neutron Imaging System (TENIS) targeted transmitted or phantom fields rather than patient-emergent backscatter Yazdandoust et al. (2021). The physical basis for the boron dependence is the $^{10}$B(n,$\alpha$)$^{7}$Li reaction: when $^{10}$B is present in tissue, thermal neutrons are preferentially absorbed before escaping, reducing the backscattered flux in proportion to local concentration. However, to our knowledge, no prior study has proposed exploiting patient-emergent backscattered thermal neutrons as an information carrier for boron concentration measurement.

In related radiation-measurement fields, the same physical phenomenon — thermal neutron backscattering carrying compositional information — has been developed into a practical measurement technique. Backscattered thermal neutron imaging has been applied to land mine detection and explosive identification, where spatially resolved hydrogen-content contrast enables localization of concealed hydrogenous objects (Bom et al., 2005; Vanier et al., 2004). Beyond imaging, neutron backscattering has been applied directly to boron content measurement. A comparative study demonstrated that the backscattering method yields approximately five times higher sensitivity than transmission for boron in water, while the 478 keV prompt-gamma neutron activation analysis (PGNAA) signal was not detectable in the same low-flux Pu-Be/NaI(Tl) configuration El Abd (2014). In nuclear fuel facilities, a dual-detector backscatter gauge has been deployed for nondestructive boron assay of neutron-absorber concrete, with the count rate following an exponential dependence on boron content Dewberry et al. (2013). These precedents demonstrate that the backscattered signal, though small in absolute magnitude, can be extracted with suitable detector and analysis strategies. Meanwhile, validated BNCT Monte Carlo workflows have demonstrated accurate neutron and dose calculations in phantom and anthropomorphic voxel geometries Hu et al. (2021). A recent proof-of-concept study of fast neutron backscatter imaging for planetary exploration demonstrated that Monte Carlo simulation can reliably reproduce experimentally observed backscatter contrast Ölçek et al. (2026). These developments motivate the present study.

We hypothesize that the backscattered thermal neutron field carries sufficient $^{10}$B-dependent information to serve as a viable measurement signal for boron concentration during the delivery of BNCT. We test this hypothesis through a proof-of-concept Monte Carlo feasibility study focusing on the following four aspects: (i) concentration and depth sensitivity in parametric homogeneous-phantom studies, (ii) diffusion-limited spatial resolution through edge-response analysis, (iii) clinical-geometry sensitivity in a voxelized patient phantom, and (iv) dosimetric perturbation from the modeled detector. A Geant4-based Monte Carlo framework is developed to couple a thin $^{nat}$LiF-converter detector model Fang et al. (2023) with homogeneous and voxelized patient phantom geometries to evaluate the feasibility of the proposed method. Two complementary metrics — Relative Fluence Reduction (RFR) and Relative Detector Signal Reduction (RDSR) — quantify the $^{10}$B-dependent signal at the physics and detector levels, respectively.

## 2. Materials and methods

### *2.1 Backscattered neutron field and measurement concept*

The measurement geometry considered in the Monte Carlo simulation is shown in Fig. 1. A neutron-transparent detector is positioned at the BSA collimator exit, between the BSA and the patient surface, capturing both forward neutrons from the primary beam and backscattered thermal neutrons emerging from the patient (Fig. 1(b)).

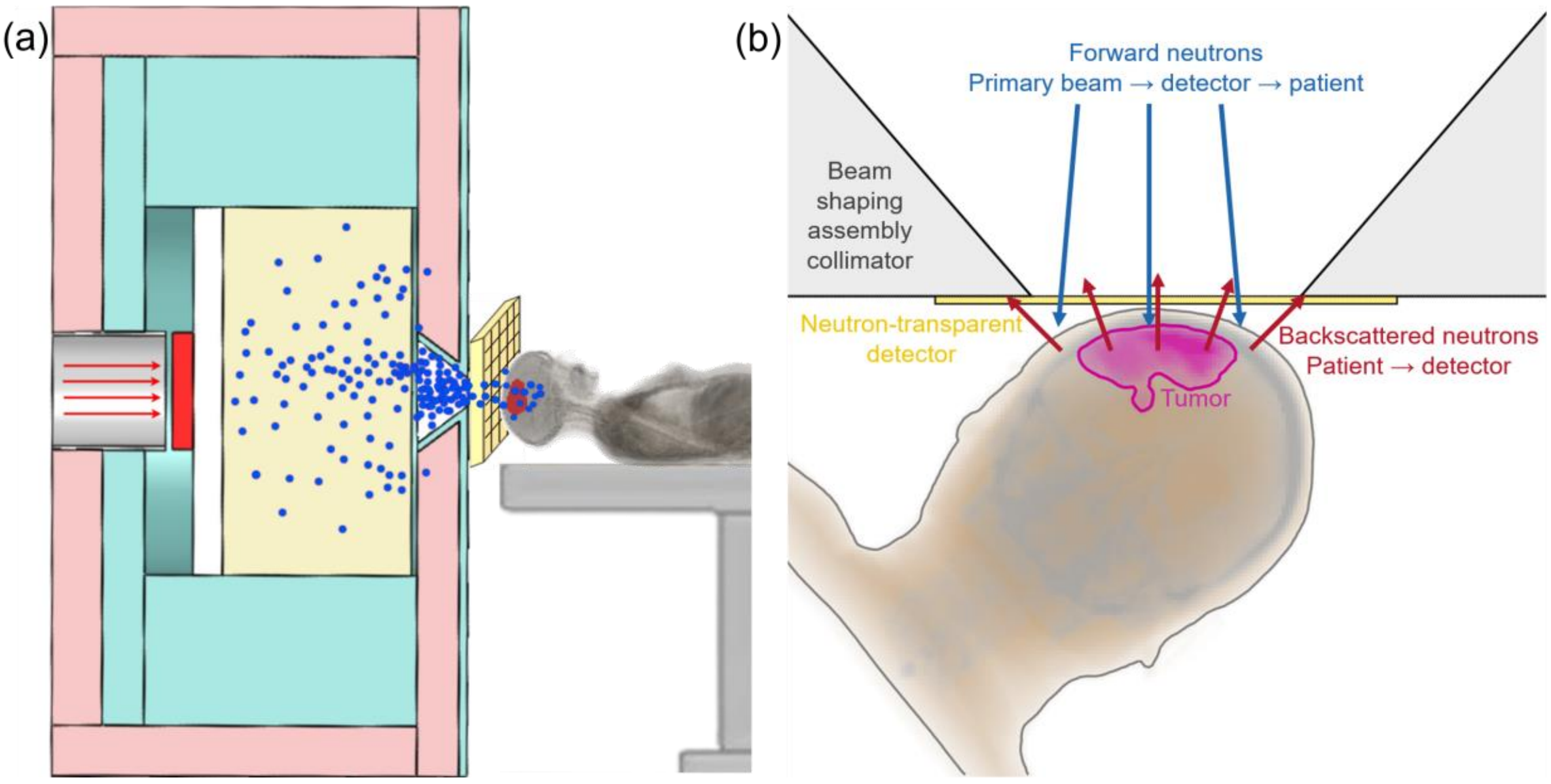

**Fig. 1.** Measurement geometry. (a) Cross-sectional overview of the beam-shaping assembly (BSA), the neutron-transparent detector, and the patient head phantom used in the Monte Carlo model. Blue dots illustrate neutron transport through the BSA; the yellow grid at the collimator exit represents the detector plane. (b) Lateral schematic showing forward neutrons (blue arrows, BSA → detector → patient) and backscattered thermal neutrons (red arrows, patient → detector). The gross tumor volume (GTV) is shown in magenta.

Oblique projections of vector fluence from the voxelized patient phantom simulation (described in Section 2.5) are shown in Fig. 2 for three energy bands — thermal ($E < 0.5$ eV), epithermal (0.5 eV $\leq E \leq$ 10 keV), and fast ($E > 10$ keV) IAEA (2023) — superimposed on semitransparent organ segmentation overlays, with arrow lengths scaled by $\sqrt{\Phi}$ ($\Phi$ means neutron fluence) to compress the dynamic range. The epithermal component is the strongest forward-directed part of the incident field and moderates to thermal energies within tissue. Thermal neutrons exhibit a different transport regime: repeated scattering in deep tissue produces near-isotropic angular distributions, a regime governed by thermal neutron diffusion Beckurts and Wirtz (1964). Near the patient surface, escaping thermal neutrons are not replenished by an equivalent inward population, producing a net outward current — the backscattered field exploited in this work. Fast neutrons contribute little to the backscatter signal. The epithermal-dominated incident beam moderates within tissue, and the resulting thermal neutrons diffuse outward, forming a backscattered field that carries spatially resolved $^{10}$B absorption information.

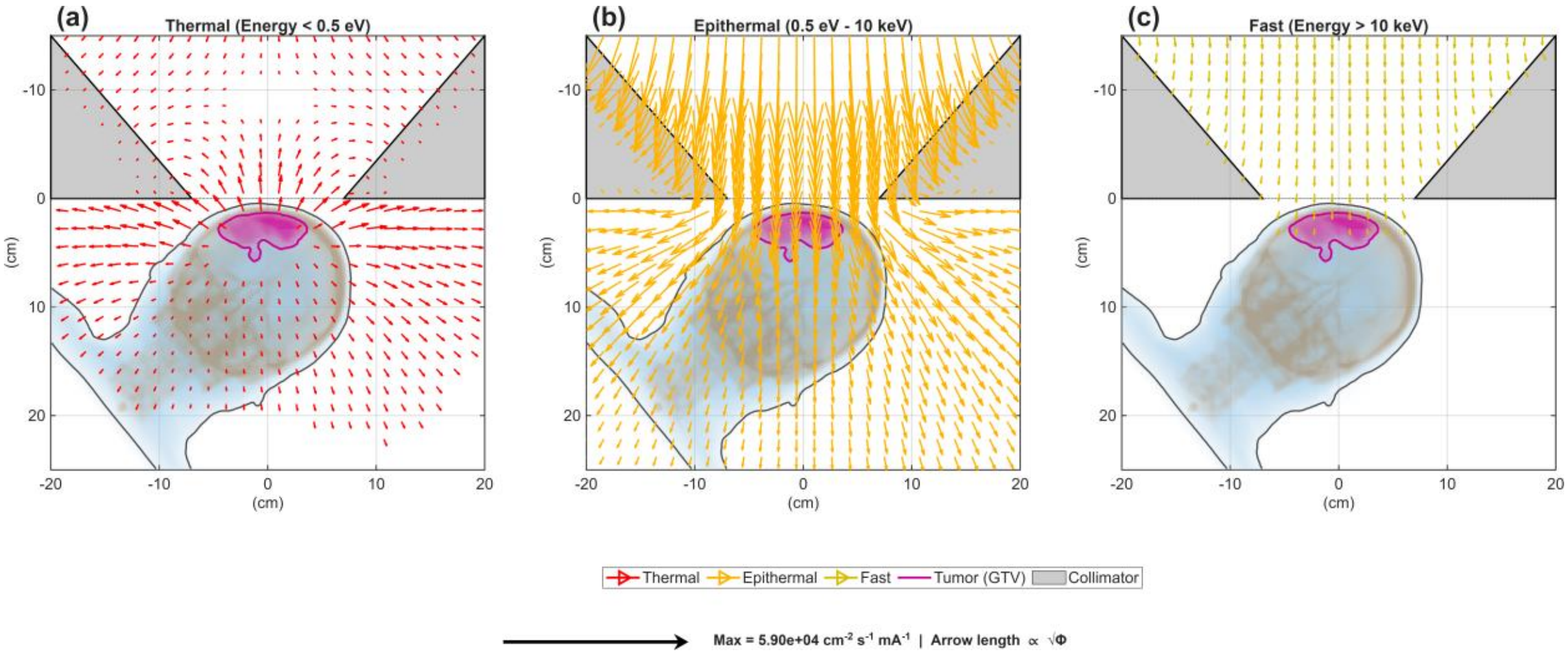

**Fig. 2.** Patient-level neutron vector fluence by energy band, displayed as oblique-view projections. (a) Thermal neutrons ($E < 0.5$ eV, red), (b) epithermal neutrons (0.5 eV $\leq E \leq$ 10 keV, orange), and (c) fast neutrons ($E >$ 10 keV, yellow). Organ segmentations are shown as semitransparent fills; the gross tumor volume (GTV) is highlighted in magenta. Arrow lengths are proportional to $\sqrt{\Phi}$; the shared normalization maximum is annotated below.

The measurement concept is illustrated in Fig. 3. The detector location shown in Fig. 1 captures both the forward component — originating from the BSA source and invariant to patient boron loading — and the backscattered component, whose intensity is modulated by $^{10}$B(n,$\alpha$)$^{7}$Li absorption in the patient. Differential imaging is defined operationally as $\Delta S(x,y) = S_{\text{baseline}}(x,y) - S_{\text{treatment}}(x,y)$, where $S$ denotes either the backscattered fluence $\Phi$ used for RFR (Eq. (1)) or the detector capture rate $C_{\text{det}}$ used for RDSR (Eq. (3)). Because the forward component cancels to first order in baseline-to-treatment subtraction, the boron-dependent backscatter variation is isolated as the measurement signal (Fig. 3). Because RFR and RDSR are baseline-normalized ratios, spatially uniform multiplicative factors — such as accelerator output drift and detector gain scaling — cancel to first order, whereas spatially structured changes such as setup shifts are preserved in the differential signal.

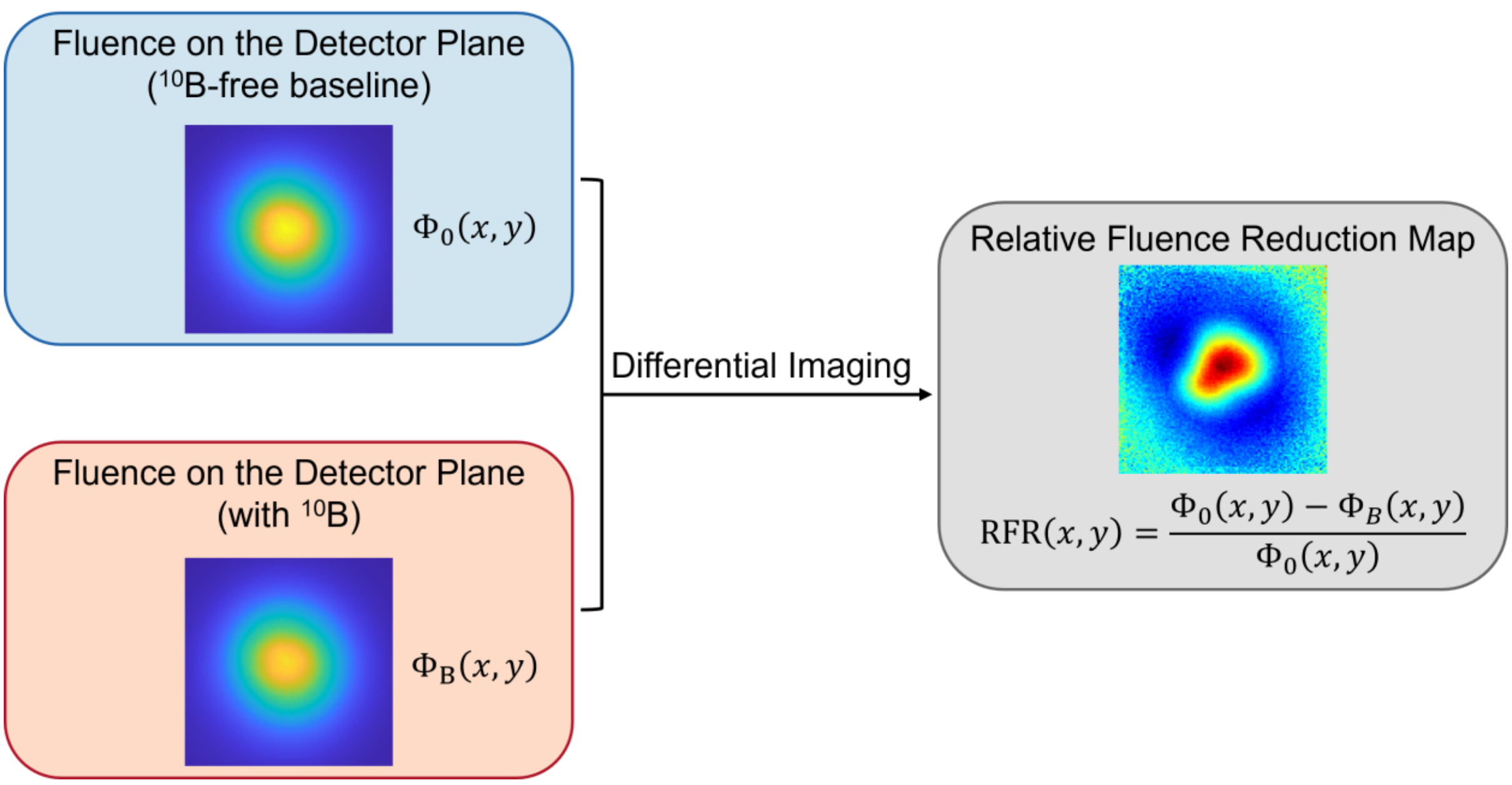

**Fig. 3.** Differential imaging workflow. Detector-plane fluence maps are scored for a $^{10}$B-free baseline, $\Phi_0(x, y)$ (upper left), and for the treatment case with $^{10}$B present, $\Phi_B(x, y)$ (lower left). Baseline subtraction yields the relative fluence reduction (RFR) map (right), defined pointwise as shown. The false-color maps are representative Monte Carlo results; the localized hot spot in the RFR map corresponds to the tumor projection. Response weighting by the $^{6}$Li capture probability converts the fluence-level RFR to the detector-level Relative Detector Signal Reduction (RDSR) defined in Eq. (3).

Two metrics were used to quantify the $^{10}$B-dependent signal. RFR measures the fractional backscattered-flux decrease; RDSR measures the fractional detector capture rate decrease. Both are defined formally below. Feasibility is assessed using two complementary phantom geometries: a homogeneous cuboid phantom for parametric characterization, and a voxelized patient phantom for clinical-geometry evaluation. Detailed configurations are described in Sections 2.2 and 2.5.

The following reporting criteria are physically motivated reference points, not clinical or regulatory requirements.

*Minimum detectable concentration difference (MDC).* The MDC under region-of-interest (ROI)-integrated analysis is characterized as a function of integration time at two levels: (i) the counting-statistics MDC, representing the Poisson limit achievable when the baseline reference has negligible uncertainty, and (ii) the baseline-uncertainty floor, representing the systematic detection limit imposed by finite baseline-reference precision. As estimated in Section 3.5, detecting a 20 ppm $^{10}$B difference would require baseline-reference precision better than ±1%, a level adopted as an illustrative reference for the floor analysis.

*Temporal resolution.* ROI-level detection of the boron-dependent signal should be achievable within a timescale compatible with treatment fractions (order of seconds to minutes).

*Spatial resolution.* The diffusion-limited spatial resolution was characterized by edge-response analysis in the homogeneous phantom. The full width at half maximum (FWHM) of the line spread function (LSF) and the corresponding modulation transfer function (MTF) serve as benchmarks for the achievable spatial information content.

*Dosimetric perturbation.* Inserting the detector into the beam path should produce limited dose-volume histogram perturbation. Treatment-time increase is reported as the primary scalar metric for beam attenuation; clinical acceptability of this trade-off is not assessed in this study.

### *2.2 Monte Carlo simulation framework*

All simulations — both homogeneous cuboid phantom and voxelized patient phantom — were performed using a Geant4-based Monte Carlo simulation framework with identical physics list and cross-section library configurations. The physics list was QGSP_BIC_AllHPT in Geant4 v11.4.0 (Agostinelli et al., 2003; Allison et al., 2006; Allison et al., 2016). QGSP_BIC_AllHP uses high-precision neutron models below 20 MeV and ParticleHP treatment for protons and light ions below 200 MeV; the T suffix activates a dedicated thermal neutron elastic-scattering treatment that improves accuracy in the sub-eV energy range critical to this study. Neutron cross-section data were provided by the G4NDL4.7.1 library (Geant4 v11.4.0), which combines evaluated nuclear data (primarily ENDF/B-VII.1) with thermal scattering kernels from JEFF-3.3 and ENDF/B-VIII.0 (Chadwick et al., 2011; Thulliez et al., 2022).

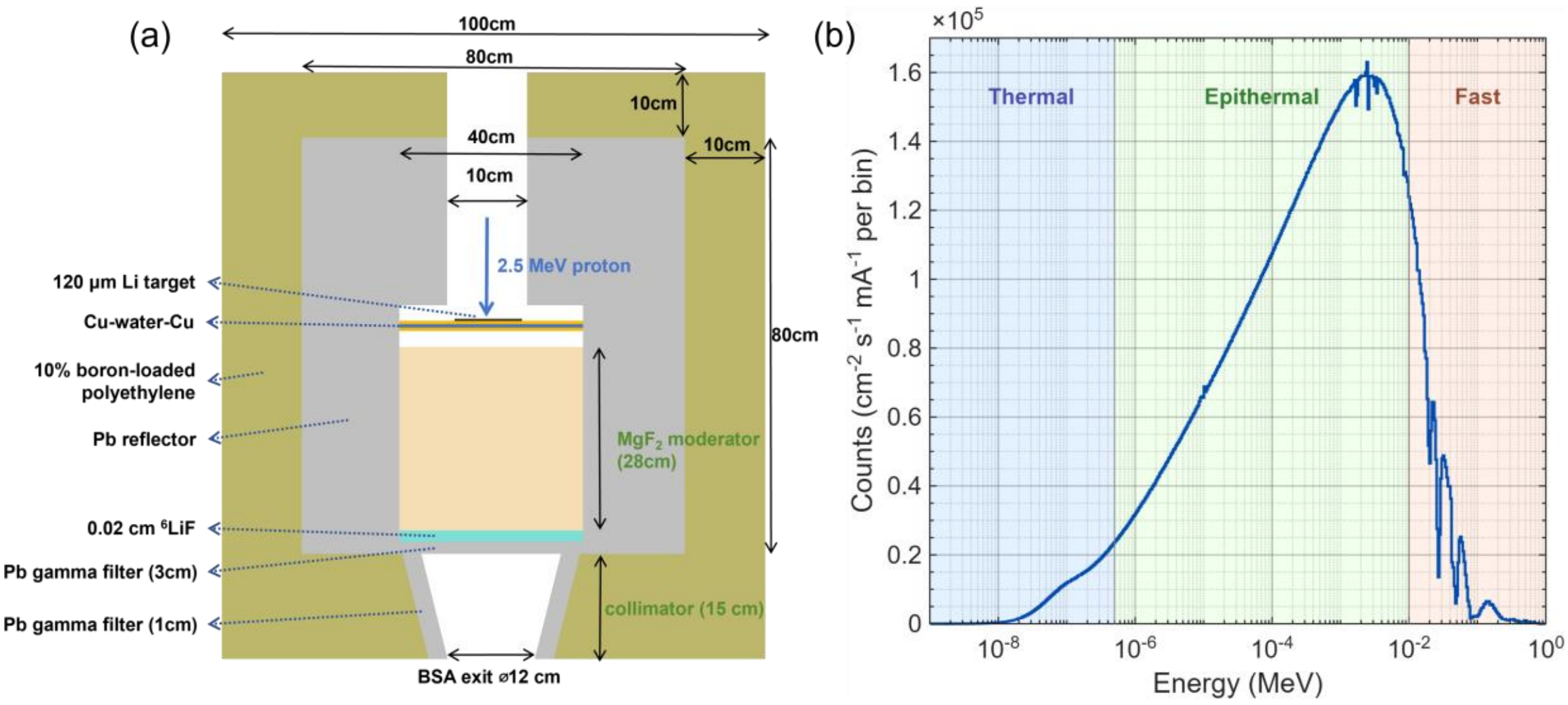

**Fig. 4.** Accelerator-based BSA configuration. (a) Internal BSA structure: 120 $\mu$m Li target, 28 cm $MgF_2$ moderator, 0.2 mm $^6$LiF filter, 3 cm Pb gamma shield, 15 cm collimator (12 cm exit diameter). (b) Neutron energy spectrum at the BSA collimator exit; shaded bands indicate thermal, epithermal, and fast energy regions.

The neutron source was modeled after an AB-BNCT BSA (Fig. 4(a)): a 2.5 MeV proton beam at 20 mA strikes a lithium target, with the resulting neutrons moderated and filtered to produce an epithermal-dominated beam exiting through a 12 cm-diameter collimator. The epithermal neutron flux at the collimator exit reaches approximately $5.2 \times 10^8$ n/cm$^2$/s, exceeding the $\geq 5 \times 10^8$ cm$^{-2}$s$^{-1}$ reference neutron beam quality factor for AB-BNCT IAEA (2023). The resulting energy spectrum at the collimator exit is shown in Fig. 4(b). The detailed BSA geometry, physics model configuration, and source-term characterization are summarized in Table 1. The BSA source-term parameters are consistent with current AB-BNCT facility reference data IAEA (2023), establishing plausibility for the modeled geometry.

**Table 1.** Beam-shaping assembly parameters and Monte Carlo physics configuration.

| Category | Parameter | Value / setting |
|---|---|---|
| Source term | Proton energy | 2.5 MeV |
| Source term | Beam current | 20 mA |
| BSA geometry | Li target thickness | 120 $\mu$m |
| BSA geometry | $MgF_2$ moderator thickness | 28 cm |
| BSA geometry | $^6$LiF filter thickness | 0.2 mm |
| BSA geometry | Pb gamma shield | 3 cm |
| BSA geometry | Collimator length / exit diameter | 15 cm / 12 cm |
| Source characterization | Epithermal flux at BSA exit | ~$5.2 \times 10^8$ n/cm$^2$/s |
| Source characterization | Exit energy spectrum | Fig. 4(b) |
| MC toolkit | Geant4 version | 11.4.0 |
| Physics list | Hadronic model | QGSP_BIC_AllHPT |
| Nuclear data | Neutron cross-section library | G4NDL4.7.1 (ENDF/B-VII.1 + JEFF-3.3/ENDF/B-VIII.0 thermal kernels) |
| Particle tracking | Secondary $\gamma$, charged particles, electrons | Enabled |
| Statistics | Homogeneous phantom | $\geq 10^9$ histories/config |
| Statistics | Patient phantom | $\geq 10^{10}$ histories/config |
| Convergence | Dose tally RSE / Fluence RSE | <0.05% / <0.02% |

The simulation credibility rests on four lines of evidence: (1) *Physics-list precedent*: the QGSP_BIC_AllHPT physics list and G4NDL neutron data have been applied in BNCT-relevant geometries by independent groups, with Geant4 vs MCNPX agreement demonstrated for boron-lined neutron detectors Ende et al. (2016) and thermal neutron scattering law processing benchmarks across Geant4 and MCNP Hartling et al. (2018); (2) *Point-level detector-response comparison*: the simulated $^6$Li capture efficiency deviates by approximately 9% from the experimental value reported by Fang et al. Fang et al. (2023) under different spectral conditions — this anchors the reaction physics but does not validate the full detection chain; (3) *Monte Carlo (MC) statistical convergence*: convergence was verified by running batches of simulations with independent random seeds until the relative standard error (RSE) stabilized below 0.05% for dose tallies and 0.02% for fluence scoring; and (4) *Thermal neutron backscatter albedo benchmark*: a monoenergetic (0.025 eV) parallel-beam simulation of a water slab using the present framework yielded a saturation albedo $\beta_{\text{sat}} = 0.802 \pm 0.001$. This value is consistent with the experimental measurement of $0.80 \pm 0.024$ Mirza et al. (2006), the diffusion-theory prediction of 0.811 Beckurts and Wirtz (1964) (within 1.1%), and a prior Geant4 v9.3 result of $0.801 \pm 0.005$ Hussain et al. (2016), providing an additional benchmark for the thermal neutron backscattering transport relevant to the present measurement concept. These lines of evidence support the physical trends and relative comparisons reported in this work; they do not constitute end-to-end experimental validation of the proposed measurement concept.

Homogeneous cuboid phantom studies used a 40 × 40 × 20 $\text{cm}^3$ phantom (voxel size 1 × 1 × 1 $\text{mm}^3$) with a spherical tumor region (radius 2.5 cm). Multiple combinations of phantom materials, tumor depths, and $^{10}$B concentrations (defined as $\mu$g $^{10}$B per gram of tissue) were simulated. Each configuration recorded the neutron fluence in a three-dimensional volumetric tally, together with a surface phase-space file storing per-particle energy and two-dimensional position at the scoring plane. Simulation statistics are given in Table 1.

Simulations in this study were executed on a dual-socket AMD EPYC 9755 server with 512 GB of memory.

### *2.3 Homogeneous phantom parametric study*

A scoring plane at the BSA collimator exit records all thermal neutrons ($E < 0.5$ eV) crossing the plane and separates them by travel direction into a BSA-to-patient (forward) component and a patient-to-BSA (backscatter) component. In the $^{10}$B-free homogeneous cuboid phantom baseline, the patient-to-BSA (backscatter) component accounts for approximately 73.8% of all thermal neutrons crossing this plane, with the forward component constituting the remainder.

The physical mechanism linking this backscattered field to $^{10}$B concentration is the $^{10}\text{B}(\text{n},\alpha)^{7}\text{Li}$ capture reaction, whose cross-section follows $1/v$ energy dependence in the thermal region (Sears, 1992; Chadwick et al., 2011). When $^{10}$B is present in tissue, thermal neutrons are preferentially absorbed before they can escape the patient, reducing the backscattered thermal flux in proportion to the local $^{10}$B concentration. Higher concentration therefore produces a measurable decrease in the outward thermal signal.

To quantify this absorption, we define the RFR as Eq. (1):

$$\mathrm{RFR}(x,y) = \frac{\Phi_0(x,y) - \Phi_B(x,y)}{\Phi_0(x,y)} \times 100\% \tag{1}$$

where $\Phi_0(x,y)$ is the baseline backscattered thermal neutron fluence at position $(x,y)$ on the scoring surface (without $^{10}$B), and $\Phi_B(x,y)$ is the corresponding fluence with $^{10}$B present under identical source normalization, geometry, energy band, and directional selection. Because $\Phi_0$ and $\Phi_B$ are spatially resolved, RFR produces a two-dimensional map of $^{10}$B absorption; it can also be integrated over any region of interest (ROI) to yield a single scalar value. The backscattered energy spectrum is dominated by thermal neutrons, with epithermal and fast contributions negligible relative to the thermal backscatter peak. As described in the field overview (Fig. 2), epithermal and fast neutrons moderate rapidly in tissue; only thermalized neutrons have sufficient diffusion length (~0.5 cm mean free path in soft tissue) to migrate back to the surface. This thermal dominance creates a natural match with the $^{6}$Li-based detector response described below.

To characterize the diffusion-limited spatial resolution, an edge-response method was applied in the same 40 × 40 × 20 cm$^3$ homogeneous cuboid phantom described in Section 2.2, with the spherical tumor region replaced by a 10 mm-thick $^{10}$B slab covering the $X > 0$ half-plane, creating a step-function boundary at $X = 0$. The depth series placed the slab at 1, 2, 3, 4, and 5 cm (100 ppm); the concentration series varied $^{10}$B from 20 to 100 ppm at a fixed depth of 2 cm; the 2 cm / 100 ppm configuration was shared by both series. Each slab configuration was paired with the $^{10}$B-free baseline. Backscattered neutrons were scored at the same scoring plane used in the parametric study (the BSA collimator exit, 4 mm from the phantom entrance surface). The $Y$-integrated one-dimensional profile, binned at 2 mm resolution perpendicular to the edge, provided the edge spread function (ESF). An error-function model was fitted to the ESF; the Gaussian standard deviation $\sigma$ of the fit yielded the LSF and its FWHM = $2\sqrt{2\ln 2}\,\sigma$. The MTF was computed analytically as the Fourier transform of the Gaussian LSF. Because the slab thickness (10 mm) is a non-negligible fraction of the FWHM at 1 cm depth (~33 mm), the extracted FWHM represents a finite-thickness edge-response estimate rather than an ideal delta-function point-spread function (PSF). The slab convolution biases the FWHM estimate upward; this bias decreases as the FWHM-to-slab-thickness ratio increases with depth.

### *2.4 Detector response modeling*

The measurement concept requires a thin, planar, two-dimensional neutron-sensitive detector with sufficient neutron transparency to limit beam attenuation. This study adopts a specific $^{nat}$LiF-converter Micromegas concept Fang et al. (2023) as the modeling reference, where the superscript "nat" denotes natural lithium isotopic abundance (7.59% $^{6}$Li); however, the physical principles — $^{6}$Li capture cross-section energy selectivity and thin-converter geometry — are not architecture-specific and apply to other candidate detector technologies (e.g., $^{6}$Li-ZnS(Ag) scintillator screens with optical readout Yamamoto et al. (2022)). The detector was modeled as a planar layer stack with a total thickness of approximately 4 mm; from source side to patient side, the successive layers are: a 0.6 mm carbon fiber support, a 15 nm $^{nat}$LiF conversion layer, a 2 mm Ar:$CO_2$ drift gap, a 15 $\mu$m stainless steel micromesh, a 0.1 mm amplification gap, a 300 nm Ge electrode, and a 1 mm printed circuit board (PCB) readout board.

The capture probability, $P_{\text{capture}}(E, \mu_z)$, was obtained from a dedicated Geant4 simulation by transporting neutrons through the complete detector structure and tallying $^{6}$Li(n,$\alpha$)t captures in the conversion layer under the AB-BNCT source model. $\mu_z = |\cos\theta_z|$ is the absolute direction cosine of the incident neutron with respect to the detector surface normal. For the thin $^{nat}$LiF conversion layer, oblique incidence increases the effective path length ($d_{\text{eff}} = d/\mu_z$) and hence the capture probability. Because the $^{6}$Li cross-section follows $1/v$ energy dependence, $P_{\text{capture}}(E, \mu_z)$ acts as a natural thermal neutron filter that preferentially weights the energy band where $^{10}$B absorption concentrates the backscatter signal. The simulated thermal neutron capture efficiency of $6.0 \times 10^{-6}$ deviates by approximately 9% from the experimental value of $6.6 \times 10^{-6}$ reported by Fang et al. for a Micromegas detector of the same design Fang et al. (2023); this difference is attributable to the spectral and angular differences between the reactor beam used in that experiment and the accelerator-based source modeled here.

The response-weighted count at each pixel is defined as Eq. (2):

$$C_{\text{det}}(x,y) = \int_0^\infty \int_0^1 \Phi_{\text{pix}}(x,y,E,\mu_z) \cdot P_{\text{capture}}(E,\mu_z)\, d\mu_z\, dE \tag{2}$$

where $\Phi_{\text{pix}}(x,y,E,\mu_z)$ is the energy- and angular-differential neutron fluence at pixel position $(x,y)$ on the detector plane (forward and backscatter combined), and $P_{\text{capture}}(E,\mu_z)$ is the Monte Carlo simulated capture probability as a function of neutron energy and incidence direction cosine. Physically, $C_{\text{det}}(x,y)$ represents the expected number of $^{6}$Li(n,$\alpha$)t captures in the thin LiF conversion layer at pixel $(x,y)$, integrated over the incident neutron energy spectrum and angular distribution under the same source normalization. Under ideal linear detector response, this detector capture rate is proportional to the primary neutron-induced detector signal before electronics-level effects.

A RDSR metric — measuring the fractional decrease in detector capture rate due to patient boron loading — is then constructed analogous to RFR (Eq. (1)) as Eq. (3):

$$\text{RDSR}(x,y) = \frac{C_{\text{det},0}(x,y) - C_{\text{det},B}(x,y)}{C_{\text{det},0}(x,y)} \times 100\% \tag{3}$$

where $C_{\text{det},0}$ and $C_{\text{det},B}$ are the $^{6}$Li capture counts under $^{10}$B-free baseline and $^{10}$B-loaded conditions, respectively, with identical source normalization. RDSR thus quantifies the fractional reduction in the detector capture rate caused by $^{10}$B absorption in the patient. Although $C_{\text{det}}$ includes both forward ($C_{\text{fwd}}$) and backscattered components, the response-weighted forward count rate within the aperture ROI ($R = 6$ cm) was $1.06 \times 10^{6}$ counts/min across all 12 patient phantom configurations — invariant to within the Monte Carlo statistical precision. Because $C_{\text{fwd},B} = C_{\text{fwd},0}$, the forward term contributes zero to the RDSR numerator; all nonzero RDSR originates exclusively from backscatter-component reduction. Electronics-level parameters (gas gain, threshold, charge sharing, dead time, dark counts) were not modeled; they affect absolute count yield but not energy-selective weighting. In the modeled Micromegas concept, gamma-background rejection is supported by the micro time projection chamber ($\mu$TPC) discrimination capability. In a comparable Li-ZnS(Ag) scintillator system, gamma-ray dose deposition in the conversion layer was simulated to be approximately 1/30 of the $^{6}$Li(n,$\alpha$) signal, and the ZnS(Ag) scintillation yield for gamma rays is a further factor of $\sim$ 50 lower than for alpha particles, rendering the gamma contribution negligible Yamamoto et al. (2022). For scalar quantification, both RFR and RDSR can be spatially integrated over any region of interest. In this work, we define a standard integration domain as a circular ROI of radius 6 cm centered on the beam axis at the scoring plane. This radius corresponds to the geometric projection of the BSA collimator exit aperture (12 cm diameter; Fig. 4(a)) onto the detector surface, encompassing the region of highest incident neutron flux and backscattered signal density. All scalar RFR and RDSR values reported in Section 3 are integrated within this aperture-projection ROI unless otherwise specified; the GTV-projection ROI analysis uses the tumor-specific region.

### *2.5 Voxelized patient phantom configurations*

The anatomical model was derived from a clinical computed tomography (CT) dataset of a patient with recurrent glioblastoma treated by BNCT, provided by Neuboron Medtech Ltd. The radiation oncologist segmented the CT images into GTV, brain, skin, bone, and remaining soft tissues. The voxel dimensions were $0.67 \times 0.67 \times 1.0$ mm$^3$, and the segmented GTV volume was 80.8 cm$^3$ with a tumor center depth of approximately 2.5 cm from the skin surface. The voxelized patient geometry is illustrated in Fig. 5, which shows representative axial and coronal CT slices with the segmented GTV contour overlaid in red.

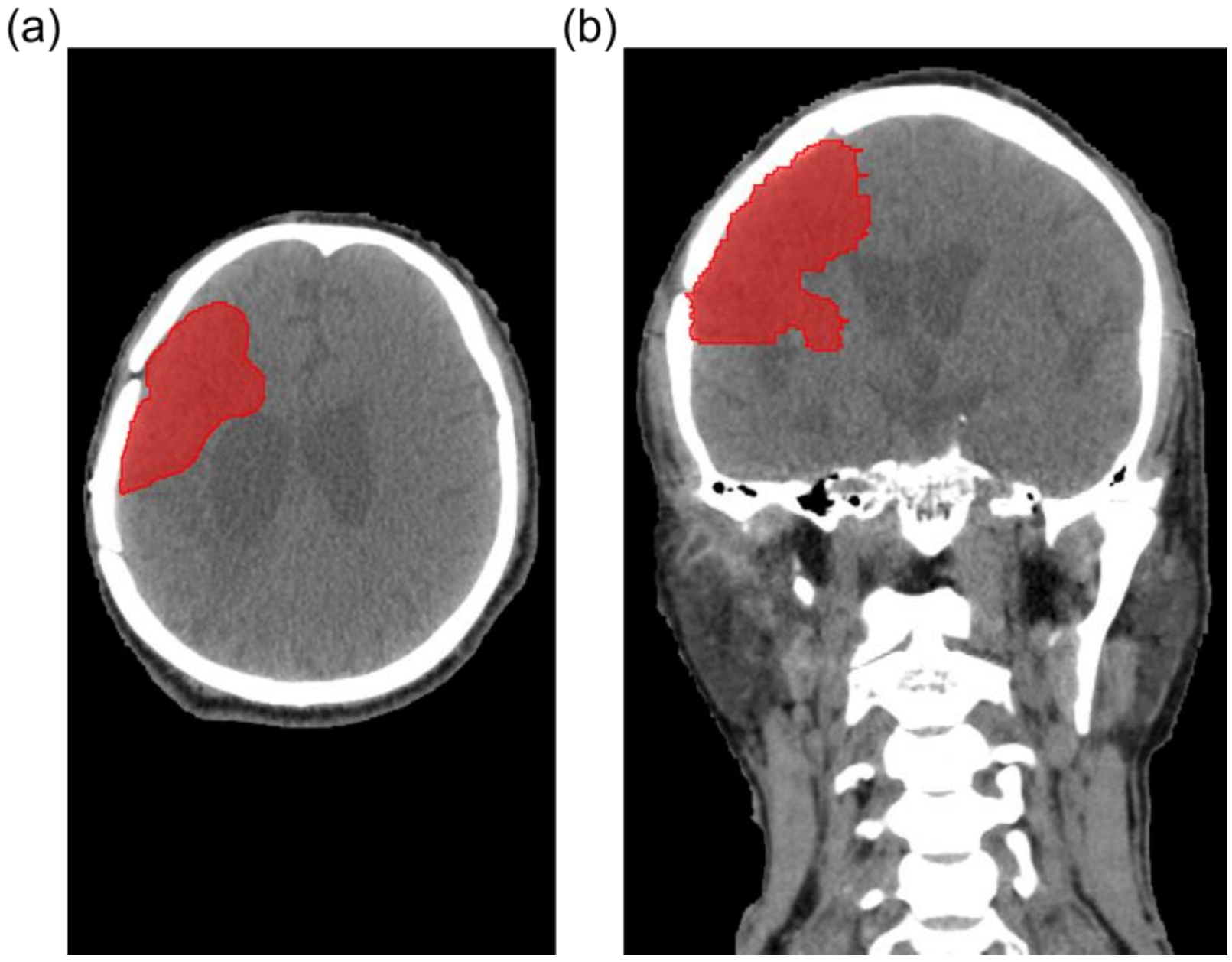

**Fig. 5.** Patient phantom derived from a clinical CT dataset of a recurrent glioblastoma case. (a) Axial and (b) coronal slices with the gross tumor volume (GTV, red overlay) indicated. The GTV (80.8 $cm^3$) is located in the right cerebral hemisphere at approximately 2.5 cm depth from the skin surface. Voxel dimensions: 0.67 × 0.67 × 1.0 $mm^3$.

Dose components — boron, nitrogen, hydrogen, and gamma — were scored as energy deposition tallies in each voxel, enabling the dosimetric comparison described below. Simulation statistics and convergence criteria are given in Table 1.

Twelve $^{10}$B configurations were investigated (Table 2), grouped into three categories. Configurations 2–5 assign $^{10}$B exclusively to the GTV, and 6–12 distribute it across skin, tumor, and normal tissue. The concentration sensitivity analysis draws on both series; the dosimetric comparison uses the reference configuration. Configuration 1 ($^{10}$B-free patient phantom) served as the computational baseline for all patient-phantom RDSR calculations; no clinical baseline-acquisition protocol was modeled.

**Table 2.** Summary of the 12 voxelized patient phantom Monte Carlo configurations grouped into three categories. Boron-10 concentrations are given in $\mu$g/g (ppm) for each tissue class. Voxelized patient phantom simulations used over $10^{10}$ total particle histories.

| # | Configuration | Skin (ppm) | Tumor (ppm) | Tissue (ppm) | Category |
|---|---|---|---|---|---|
| 1 | Baseline ($^{10}$B-free) | 0 | 0 | 0 | Baseline |
| 2 | Tumor-only 20 ppm | 0 | 20 | 0 | Tumor-only |
| 3 | Tumor-only 40 ppm | 0 | 40 | 0 | Tumor-only |
| 4 | Tumor-only 60 ppm | 0 | 60 | 0 | Tumor-only |
| 5 | Tumor-only 80 ppm | 0 | 80 | 0 | Tumor-only |
| 6 | Clinical low-background | 10 | 60 | 10 | Clinical |
| 7 | Clinical low-tumor | 13 | 30 | 10 | Clinical |
| 8 | Clinical reference 20 ppm | 26 | 20 | 20 | Clinical |
| 9 | Clinical reference 40 ppm | 26 | 40 | 20 | Clinical |
| 10 | Clinical reference 60 ppm | 26 | 60 | 20 | Clinical |
| 11 | Clinical reference 80 ppm | 26 | 80 | 20 | Clinical |
| 12 | Clinical reference 100 ppm | 26 | 100 | 20 | Clinical |

Abbreviations: CT, computed tomography; GTV, gross tumor volume; ppm, $\mu$g $^{10}$B per gram of tissue.

### *2.6 Dosimetric impact assessment*

To quantify the dosimetric perturbation introduced by the modeled detector, two treatment plans were compared using the same voxelized patient phantom, beam configuration (2.5 MeV, 20 mA), and boron distribution (Skin 26 ppm, Tumor 60 ppm, Tissue 20 ppm): a reference plan without the detector and a plan with the detector inserted at the collimator exit. Dose components were converted to Gy-Eq using established relative biological effectiveness (RBE) and compound biological effectiveness (CBE) factors (Coderre et al., 1993; Coderre and Morris, 1999; Hopewell et al., 2011; Kumada H, 2018). The isoeffective dose was calculated voxel-wise as Eq. (4):

$$D_{\text{Iso-E}}(\text{Gy-Eq}) = D_B \times \text{CBE}_t + D_N \times \text{RBE}_N + D_H \times \text{RBE}_H + D_\gamma \times \text{RBE}_\gamma \tag{4}$$

where $D_B$, $D_N$, $D_H$, and $D_\gamma$ are the physical dose components from boron, nitrogen, fast neutron (hydrogen recoil), and gamma interactions, respectively, and $t$ denotes the tissue class. $\mathrm{RBE}_N$ and $\mathrm{RBE}_H$ were both set to 3.2, and $\mathrm{RBE}_\gamma$ to 1.0. $\mathrm{CBE}_t$ was assigned as 3.8 for GTV, 2.5 for skin, and 1.35 for normal tissue. Bone was modeled without $^{10}$B uptake.

Both plans were normalized to the same prescription constraint: the maximum normal-brain dose was set to 12 Gy-Eq Miyatake et al. (2016) at 20 mA beam current.

Dose-volume histograms (DVHs) were compared for GTV, brain, skin, and other tissues, together with two-dimensional weighted-dose distributions in the coronal plane. Treatment time — defined as the irradiation duration required to reach the prescription constraint — served as the primary scalar metric for beam attenuation.

## 3. Results and discussion

### *3.1 Homogeneous phantom characterization*

Simulations of the homogeneous cuboid phantom characterize the spatial and spectral properties of the backscattered neutron field and its sensitivity to $^{10}$B concentration and tumor depth. The backscatter component dominates the outward thermal field (~73.8% at baseline). Fig. 6(a,b) show the neutron fluence maps relatively produced by the epithermal-dominant source beam and the thermal backscatter field at the phantom entrance surface.

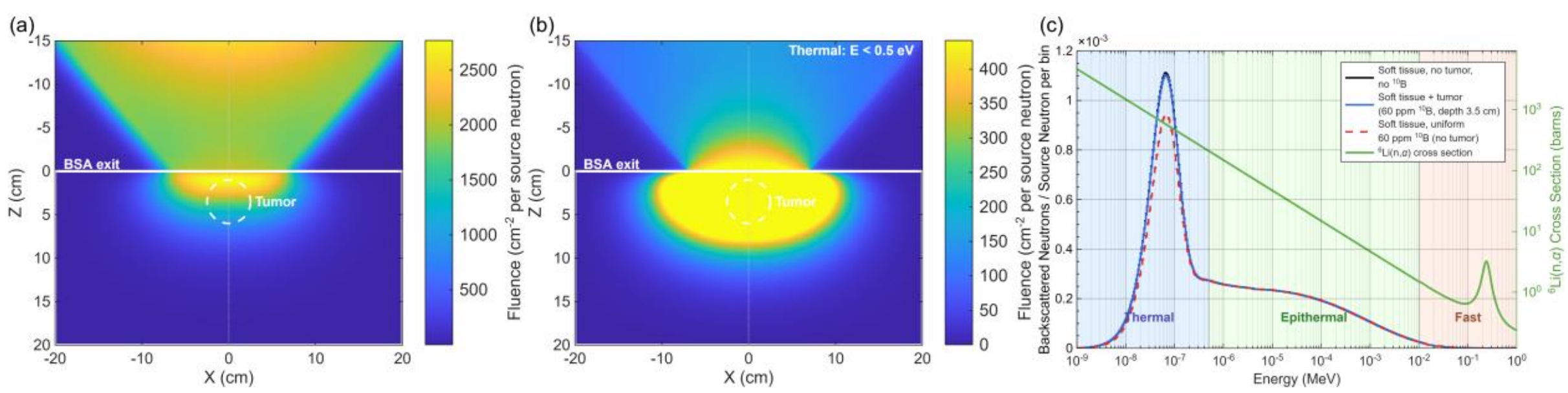

**Fig. 6.** Neutron field and backscattered energy spectrum in the homogeneous cuboid phantom. (a) Total neutron fluence in the central X–Z cross-section ($\mathrm{cm}^{-2}$ per source neutron). The dashed circle marks the spherical tumor region (radius 2.5 cm, depth 3.5 cm, 60 ppm $^{10}$B). (b) Thermal neutron fluence ($E < 0.5$ eV) in the same cross-section. (c) Backscattered energy spectrum for three configurations: soft-tissue baseline (black), localized

60 ppm tumor (blue), and uniform 60 ppm (red dashed). Right axis (green): $^{6}$Li(n,$\alpha$)t capture cross-section. Shaded bands indicate thermal, epithermal, and fast energy regions.

The energy spectrum of backscattered neutrons (Fig. 6(c)) confirms that the signal is concentrated in the thermal energy band. Three representative configurations are compared: a $^{10}$B-free soft-tissue baseline (black), a localized tumor at 3.5 cm depth with 60 ppm $^{10}$B (blue), and a uniform 60 ppm distribution throughout the entire phantom volume (red dashed). Differences between configurations appear exclusively at the thermal peak (~0.025 eV): the localized $^{10}$B case shows selective absorption relative to baseline, and the uniform case — representing an upper bound on absorption — produces a more pronounced reduction.

Parametric dependence is characterized through concentration and depth series. Concentration-series spectra (Fig. 7(a)) show monotonically decreasing thermal peak height with approximately equal spacing. The depth series (Fig. 7(b)) shows signal decay with depth, converging toward the baseline beyond ~7 cm.

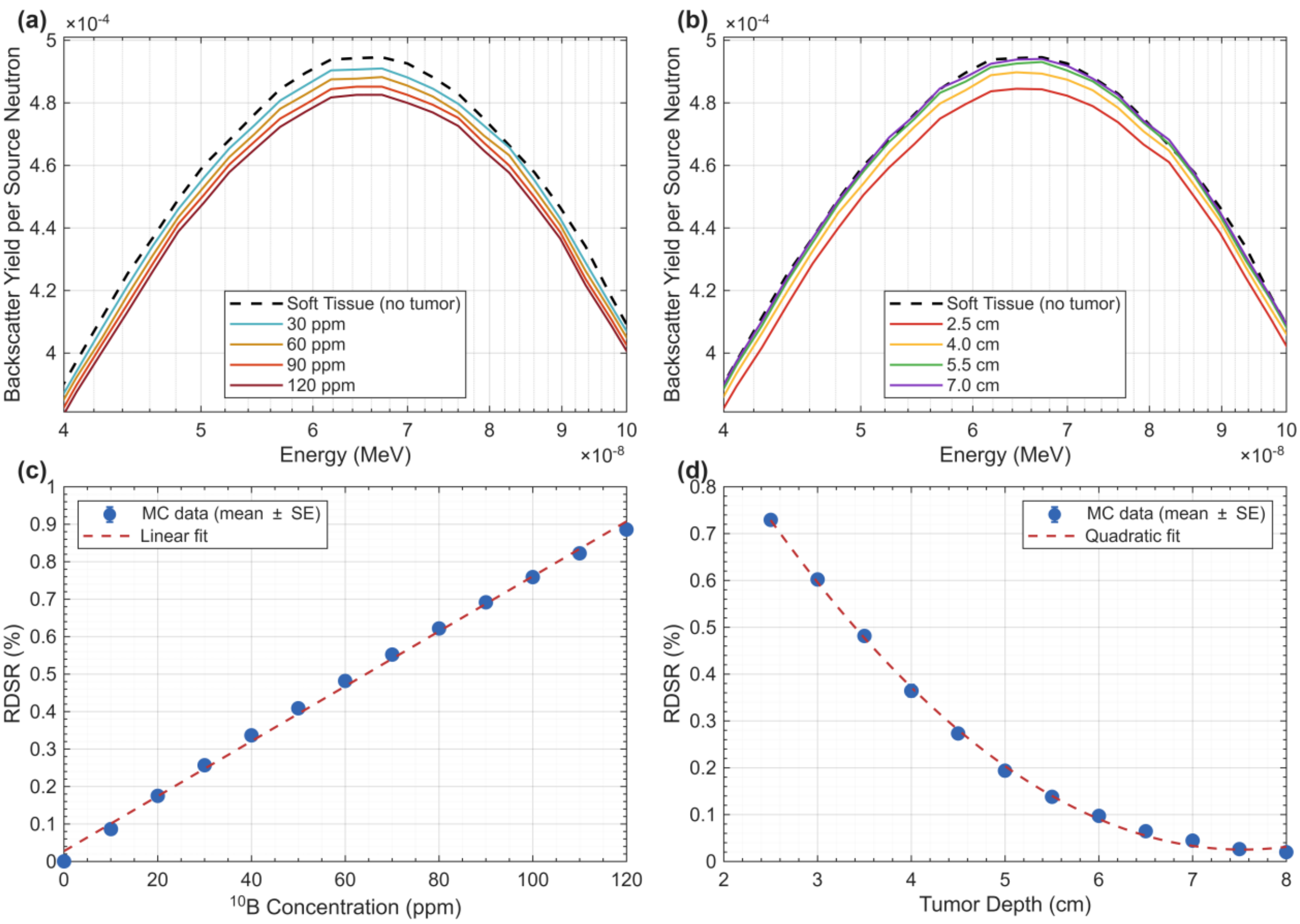

**Fig. 7.** Parametric dependence of backscattered spectra and RDSR in the homogeneous cuboid phantom, with all RDSR values integrated within the BSA aperture-projection ROI ($R = 6$ cm). (a) Thermal-peak spectra for concentration series at 3.5 cm depth: baseline, 30, 60, 90, and 120 ppm. (b) Depth series at 60 ppm: baseline and tumor depths 2.5, 4.0, 5.5, and 7.0 cm. (c) RDSR versus $^{10}$B concentration with linear fit ($y = 0.00733x + 0.0275,\ R^2 = 0.9974$). (d) RDSR versus tumor depth with quadratic fit ($y = 0.0278x^2 - 0.418x + 1.60$, $R^2 = 0.9990$). Error bars represent mean ± standard error.

Quantitative RDSR fits consolidate the spectral observations. RDSR versus concentration (Fig. 7(c)) is linear ($R^2 = 0.997$), with sensitivity ~0.0073%/ppm at 3.5 cm depth. RDSR versus tumor depth (Fig. 7(d)) is well described by a second-order polynomial ($R^2 = 0.999$), declining from 0.73% at 2.5 cm to 0.020% at 8.0 cm. Beyond ~6 cm, RDSR decays below 0.1%, a level where the backscatter contribution becomes negligible relative to practical counting-statistics uncertainty; the polynomial fit is valid only within the tested 2.5–8.0 cm range.

### *3.2 Sensitivity of patient-specific $^{10}$B concentration*

The same analysis framework was applied to the voxelized patient phantom with realistic $^{10}$B distributions across skin, tumor, and normal tissue. Fig. 8 presents the three-dimensional context, while Fig. 9 presents two-dimensional RDSR maps and qualitative one-dimensional profiles.

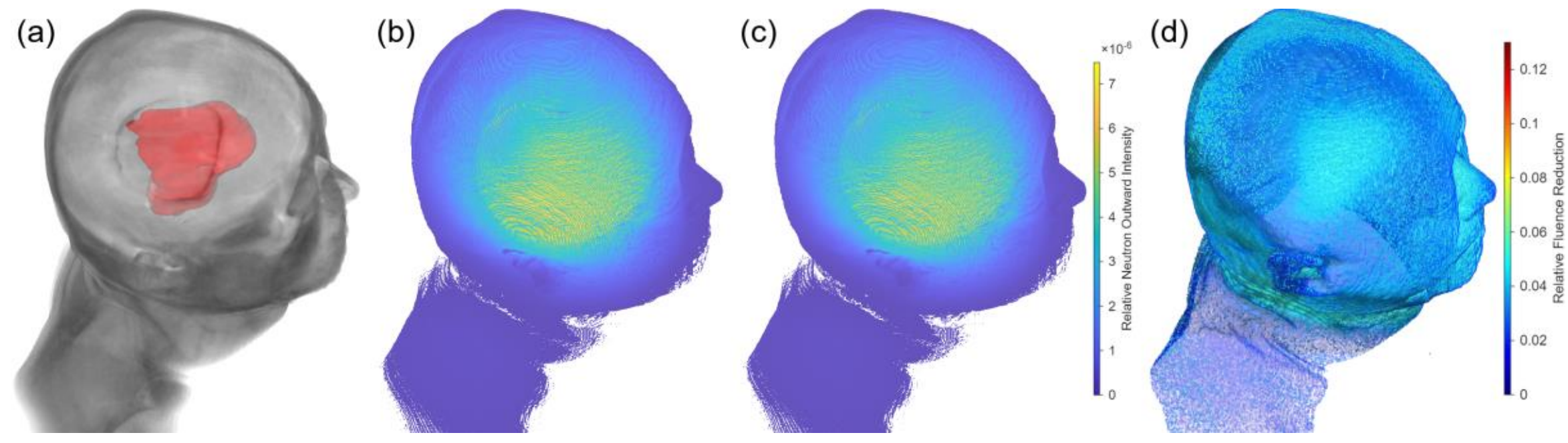

**Fig. 8.** Three-dimensional context of the patient-specific backscattered neutron field. (a) CT volume rendering with the GTV segmentation highlighted (3D Slicer v5.8.1). (b) Baseline (Configuration 1) outward neutron intensity maximum intensity projection (MIP). (c) Clinical

reference (Configuration 10) outward neutron intensity MIP. (d) RFR volume rendering (dimensionless color scale, 0–0.13).

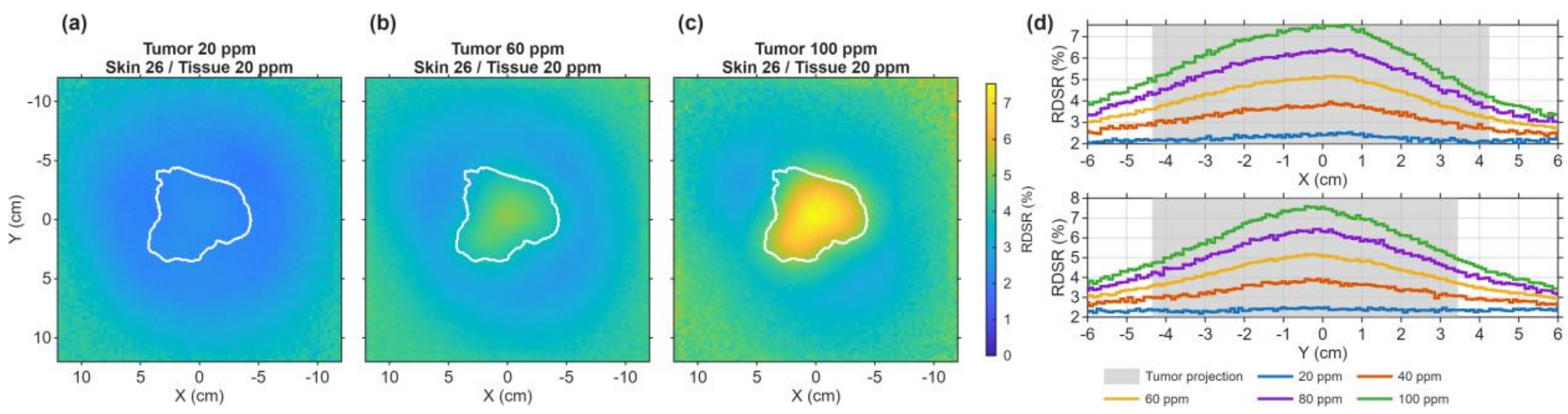

**Fig. 9.** Two-dimensional RDSR maps and one-dimensional profiles in the patient geometry. (a–c) Pixel-wise RDSR maps for tumor $^{10}$B concentrations of 20, 60, and 100 ppm (Configurations 8, 10, and 12; shared color scale), with the projected GTV contour overlaid as a white solid outline. RDSR maps are displayed after 3 mm mean-binning resampling from the original 0.5 mm scoring resolution. (d) One-dimensional RDSR profiles along horizontal and vertical axes through the GTV center. The profiles are intended for qualitative spatial comparison; quantitative concentration response is reported using ROI-integrated metrics (Fig. 10).

Fig. 8 establishes the three-dimensional context. The relative outward neutron intensity MIP of the clinical reference (Fig. 8(c)) is visually indistinguishable from the baseline (Fig. 8(b)); the RFR volume rendering (Fig. 8(d)) reveals the $^{10}$B-dependent signal as a diffuse pattern across the body surface.

A GTV-localized signal intensifying from 20 to 100 ppm is superimposed on a background floor (~3–5%) from distributed tissue $^{10}$B (Fig. 9(a–c)). The GTV ROI integral analysis (Fig. 10(a)) separates the total response-weighted signal reduction into tumor and background $^{10}$B contributions. The background contribution remains approximately constant at 1.1–1.3% across the concentration range, while the tumor contribution increases from 1.4% at 20 ppm to 5.3% at 80 ppm, where it dominates the background at a ratio of approximately 3.8:1. Although the background $^{10}$B introduces a baseline shift, it does not reduce the sensitivity to tumor concentration changes — the tumor-only and clinical series exhibit consistent concentration-response trends across the tested range.

Signal-to-noise ratio (SNR), defined as $\mathrm{RDSR_{GTV}} \times \sqrt{N_{\mathrm{baseline,GTV}}}$ where $N_{\mathrm{baseline,GTV}}$ is the GTV ROI baseline count accumulated over integration time $t$, ranges from 5.30 (20 ppm) to 14.00 (100 ppm) at $t = 1$ s. For the clinical reference (60 ppm), SNR reaches 9.89 at 1 s and scales as $\sqrt{t}$ with longer integration.

The patient geometry yields higher RDSR sensitivity than the homogeneous phantom at comparable boron concentrations — the patient-specific aperture RDSR slope (0.036 %/ppm, Configurations 8–12) exceeds the homogeneous phantom slope (0.0073 %/ppm, tumor-only series) by a factor of approximately 4.9. Both geometries use the same BSA aperture-projection ROI ($R = 6$ cm). The sensitivity difference is attributable to anatomical factors (shapes and materials). The layered heterogeneous anatomy of the head — bone mineral composition, brain/bone/skin/tumor interfaces, and asymmetric tissue boundaries — modifies neutron scattering, moderation, and surface leakage pathways in ways that a uniform water-equivalent slab does not capture. This enhancement cannot be predicted from simplified geometries, which highlights the need for patient-specific Monte Carlo simulation to characterize the expected signal for each individual anatomy rather than relying on universal calibration factors derived from a single patient or phantom.

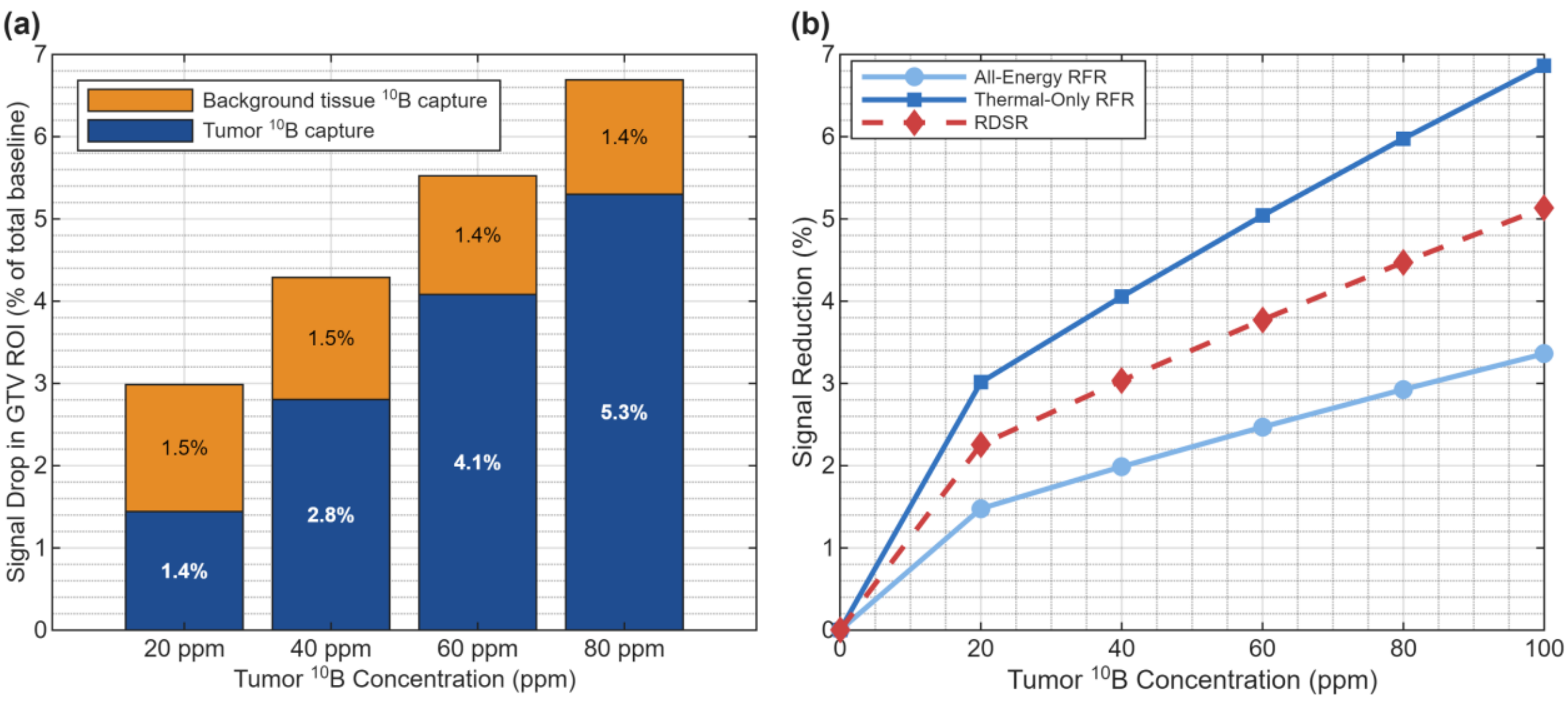

**Fig. 10.** Signal decomposition and energy-selective response in the patient geometry. (a)

Stacked bar chart of the GTV ROI response-weighted signal drop decomposed into tumor $^{10}$B (blue) and background tissue $^{10}$B (orange) contributions at four tumor concentrations (20–80 ppm, all with Skin 26 / Tissue 20 ppm). Percentage annotations show each component. (b) Three metrics integrated within the BSA aperture-projection ROI ($R = 6$ cm) versus tumor concentration (20–100 ppm): all-energy backscatter RFR (circles), thermal-only backscatter RFR (squares), and total-signal RDSR (diamonds, dashed). All-energy and thermal-only RFR use only the backscatter direction; RDSR includes the forward component in its denominator.

### *3.3 Signal decomposition and energy-selective response*

This section examines how the $^{6}$Li detector response enhances the boron-dependent signal relative to unweighted fluence metrics. The $^{6}$Li capture cross-section provides intrinsic energy selectivity in the detector response: it is two to three orders of magnitude higher at thermal energies than at epithermal energies. Three metrics — all-energy RFR, thermal-only RFR, and total-signal RDSR — represent successive levels of energy and direction selection.

Three metrics (Fig. 10(b)), all integrated within the BSA aperture-projection ROI ($R = 6$ cm), illustrate the different signal extraction levels. Their ordering — all-energy RFR < RDSR < thermal-only RFR — reflects two competing effects: restricting to thermal neutrons removes insensitive epithermal and fast backscatter (increasing the metric), while the total-signal RDSR denominator includes the forward component that does not carry boron information (decreasing the metric relative to backscatter-only thermal RFR). At 60 ppm (clinical reference), the RDSR is 3.77%, positioned between the all-energy RFR (2.47%) and thermal-only RFR (5.04%). The $^{6}$Li cross-section provides intrinsic energy selectivity that preferentially weights the thermal energy band where $^{10}$B absorption is concentrated, without requiring external energy discrimination.

This energy selectivity is a physics-level property of the $^{6}$Li detection mechanism, independent of downstream gas gain, electronic threshold, or readout architecture. It has two effects: preferential weighting of the thermal band where $^{10}$B absorption concentrates the signal, and suppression of the forward component from ~26% (unweighted flux) to ~17% (response-weighted signal). Gamma-induced events may reduce signal-to-noise without altering the energy-selective mechanism. Forward-beam optical imaging methods — whether $^{6}$Li-based Yamamoto et al. (2022) or $^{10}$B-based Maeda et al. (2024) — share the same $1/v$ thermal neutron selectivity exploited here.

### *3.4 Diffusion-limited spatial resolution: homogeneous-phantom characterization*

The spatial response of the backscattered neutron field was characterized using the edge-response method described in the Methods, applied to the homogeneous cuboid phantom. Because these measurements were performed in a uniform medium with a planar geometry, the resulting FWHM values represent homogeneous-phantom edge-response widths rather than the spatial resolution achievable in a heterogeneous patient anatomy. Nine unique $^{10}$B slab configurations plus a boron-free baseline were simulated: five depth points at 100 ppm and five concentration points at 2 cm depth, with the 2 cm / 100 ppm configuration shared by both series.

Fig. 11(a) shows the edge-response geometry and a representative RFR map at $d = 2$ cm. Fig. 11(b) presents the response-weighted ESF for the depth series; the FWHM of the corresponding Gaussian LSF quantifies the diffusion-limited spatial resolution. The FWHM increases monotonically with slab depth: 32 mm at 1 cm, 57 mm at 2 cm, 83 mm at 3 cm, 113 mm at 4 cm, and 176 mm at 5 cm (Fig. 11(c)). The MTF = 0.5 cutoff spatial frequency, derived analytically from the Gaussian $\sigma$, decreases from 0.14 cycles/cm at 1 cm depth to 0.04 cycles/cm at 4 cm depth, corresponding to half-period resolvable feature widths of approximately 3.6 cm and 12.6 cm, respectively; at 5 cm depth, the cutoff decreases further to 0.025 cycles/cm (half-period ~20 cm), indicating that useful spatial modulation is essentially lost at this depth. Even at the shallowest depth investigated, the spatial resolution remains limited to centimeter-scale structures.

The FWHM exhibits weak concentration dependence: at 2 cm depth, the response-weighted FWHM decreases gradually from 61 mm at 20 ppm to 57 mm at 100 ppm, with the decrease concentrated between 20 and 40 ppm and stabilizing above ~60 ppm (Fig. 11(c)). The total variation across the 20–100 ppm range is less than 7%, compared with the 5.5-fold increase over the 1–5 cm depth range. This weak concentration dependence confirms that the spatial resolution is governed primarily by the thermal neutron multiple-scattering transport geometry — which determines the point-spread kernel — with $^{10}$B absorption producing only a minor sharpening of the edge response at higher concentrations. The diffusion-limited resolution can therefore be characterized once for a given depth and applied across concentration levels without recalibration.

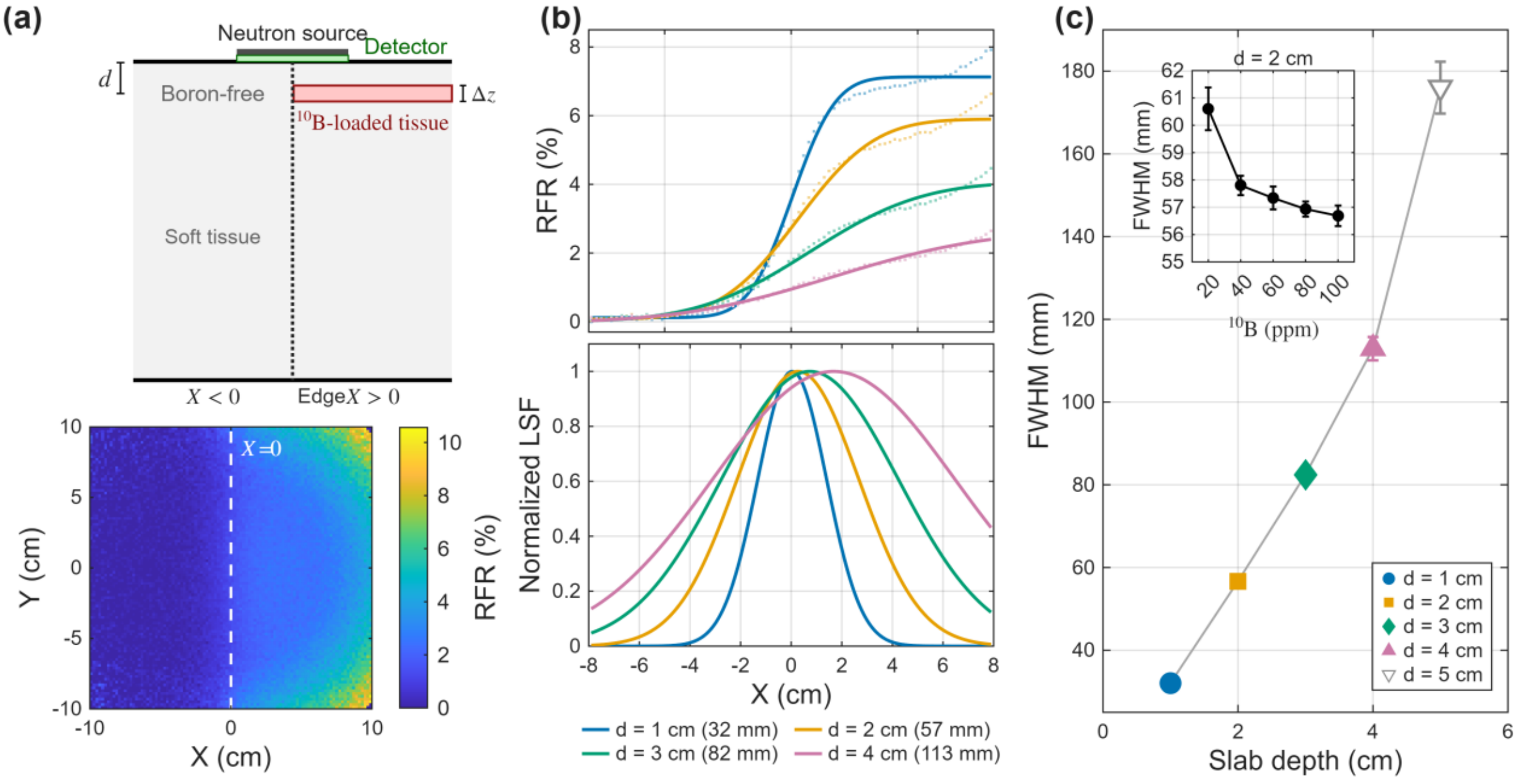

**Fig. 11.** Diffusion-limited spatial resolution of the backscattered neutron imaging concept, characterized by edge-response analysis in the homogeneous phantom. (a) Cross-sectional schematic of the edge-response geometry (upper) and two-dimensional RFR map at the scoring plane for $d = 2$ cm, 100 ppm $^{10}$B (lower). The $^{10}$B-doped slab (10 mm thick) occupies the $X > 0$ half-plane at depth $d$; the white dashed line marks the edge boundary at $X = 0$. (b) Response-weighted ESF (upper; dots: MC data, lines: error-function fits) and normalized Gaussian LSF (lower) for slab depths of 1–4 cm (100 ppm $^{10}$B). (c) Response-weighted FWHM versus slab depth (main plot, $d = 1$–5 cm, 100 ppm) and versus $^{10}$B concentration at $d = 2$ cm (inset; FWHM decreases gradually from 61 mm at 20 ppm to 57 mm at 100 ppm, stabilizing above ~60 ppm). Error bars represent bootstrap standard deviations (1000 resamples).

These FWHM values quantify the diffusion-limited blurring in a homogeneous medium,

establishing that the backscattered neutron field carries centimeter-scale spatial information that degrades monotonically with depth over the 1–5 cm range. In a heterogeneous patient anatomy, tissue interfaces, bone structures, and irregular surfaces would modify the effective point-spread function in ways not captured by the homogeneous model; the reported FWHM values should therefore be interpreted as baseline estimates of the diffusion-limited spatial scale rather than as system-level resolution specifications. The measurement concept is better suited to coarse spatial mapping and ROI-level trend measurement (Section 3.5) than to sub-centimeter boron imaging.

### *3.5 ROI-level detectability and performance metrics*

The measurement signal supports two complementary readout modes: full-field two-dimensional mapping for spatial characterization, and ROI-integrated scalar tracking for high-statistics trend detection. The following analysis focuses on the ROI-integrated mode.

To assess practical detectability, count rates were estimated under modeled ideal detector conditions at 20 mA beam current. All count rates are reported within the BSA aperture-projection ROI ($R = 6$ cm, area $\approx 113$ cm$^2$), consistent with the spatial domain used for RFR and RDSR analysis. Under the $^{10}$B-free baseline (Configuration 1), the total detector signal — combining forward ($^6$Li captures from the BSA beam traversing the converter layer) and backscatter ($^6$Li captures from patient-scattered neutrons) — is approximately $1.0 \times 10^5$ counts/s within the aperture ROI, corresponding to approximately $9.1 \times 10^2$ counts/s/cm$^2$.

ROI-level integration provides a statistical advantage over spatially resolved imaging. For clinically representative configurations (tumor 60–100 ppm), the aperture-ROI RDSR — the area-integrated form of Eq. (3) over the BSA aperture ROI — is on the order of 3.8–5.1%. Assuming that the baseline reference is acquired with negligible statistical uncertainty (e.g., from a high-statistics simulation or a prolonged pre-treatment calibration exposure), Poisson statistics yield a $3\sigma$ detection time (Eq. (5)):

$$t_{3\sigma} = \frac{9}{\mathrm{RDSR}^2 \times C_{\mathrm{ROI}}} \tag{5}$$

where $C_{\mathrm{ROI}}$ is the baseline ROI count rate (counts/s). In Eq. (5), RDSR is expressed in fractional form (e.g., 0.038 for 3.8%). For RDSR = 3.8% (Configuration 10, tumor 60 ppm), this gives $t_{3\sigma} \approx 0.06$ s; for RDSR = 5.1% (Configuration 12, tumor 100 ppm), $t_{3\sigma} \approx 0.03$ s. If both baseline and treatment images are acquired with comparable counting statistics, these estimates would increase by approximately a factor of two. In either case, ROI-level concentration-change detectability falls within a sub-second timescale as a Poisson counting-statistics lower bound under idealized modeled conditions — specifically, perfect baseline knowledge, ideal detector response (no dead time, pile-up, or gain nonuniformity), and absence of gamma background. In a physical implementation, detection times would exceed these estimates owing to the omitted factors.

Pixel-level integration time scales inversely with pixel area. Given the diffusion-limited FWHM of 32–176 mm (Section 3.4), sub-centimeter pixel sizes do not improve spatial information content. At 3 × 3 $\mathrm{mm}^2$ rebinning — well below the diffusion-limited resolution — the pixel-level $t_{3\sigma}$ is approximately 5.6 s at 3.8% RDSR, compared with ~0.5 s at 10 × 10 $\mathrm{mm}^2$, representing a spatial-temporal trade-off adjustable through analysis-level rebinning.

The MDC was computed from the patient-specific RDSR concentration-response slope under two ROI definitions: the BSA aperture-projection ROI ($R = 6$ cm, slope $k = 0.036$ %/ppm, baseline count rate $1.02 \times 10^5$ counts/s) and the GTV-projection ROI (48.7 $\mathrm{cm}^2$, slope $k = 0.046$ %/ppm, baseline count rate $5.31 \times 10^4$ counts/s). The GTV-projection ROI is defined from the treatment-planning CT segmentation and beam/detector geometry — not inferred from the backscatter image — representing a best-case that requires tumor geometry (GTV segmentation); the aperture ROI requires only beam-geometry information, providing a prior-independent readout. MDC was calculated as $\Delta C_{\min}(t) = \sqrt{18}/(k \cdot \sqrt{C_{\mathrm{ROI}} \cdot t})$, where $k$ is expressed in fractional form (e.g., $3.6 \times 10^{-4}$ $\mathrm{ppm}^{-1}$ for the 0.036 %/ppm aperture-ROI slope), assuming comparable baseline and treatment counting statistics.

Fig. 12 shows MDC versus integration time for both ROIs, with the analysis separated into two levels. At the counting-statistics level — assuming a baseline reference with negligible uncertainty — the MDC follows a $1/\sqrt{t}$ decay: at 60 s integration, the aperture ROI achieves a counting-statistics MDC of 4.8 ppm and the GTV ROI achieves 5.1 ppm; the 20 ppm reporting reference is crossed at 3.4 s (aperture) and 3.9 s (GTV). The aperture ROI yields slightly lower MDC, because the larger area (113 vs 49 $cm^2$) provides a count-rate advantage that compensates for its smaller slope. These values represent counting-statistics lower bounds that quantify the intrinsic information capacity of the backscattered signal: the RDSR concentration-response slope and count rate are sufficient to resolve single-digit ppm differences given adequate integration time and a perfectly known baseline.

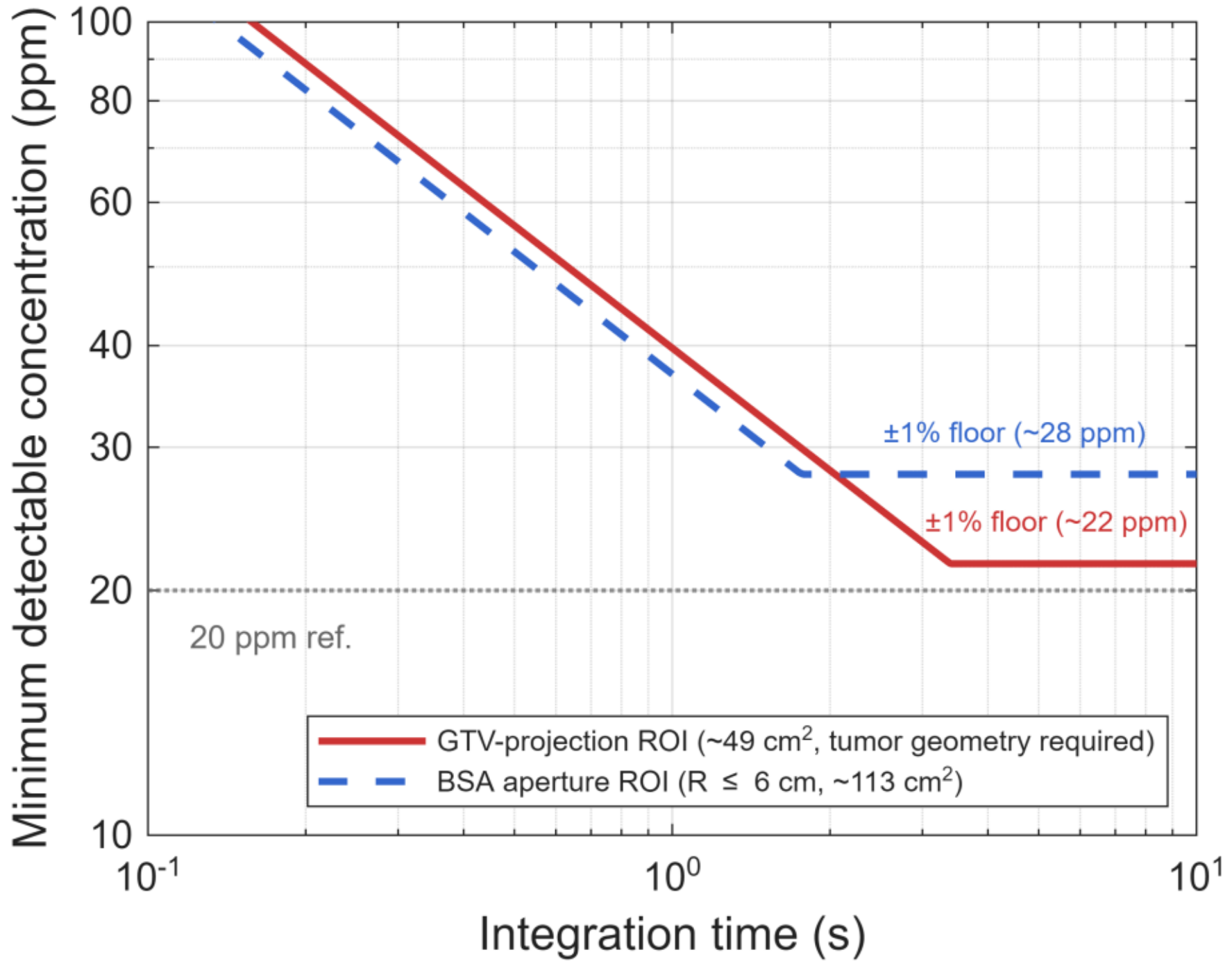

**Fig. 12.** Minimum detectable concentration difference (MDC) versus integration time for the BSA aperture-projection ROI (blue dashed) and GTV-projection ROI (red solid), computed from patient-specific RDSR concentration-response slopes (Configurations 8–12) assuming comparable baseline and treatment counting statistics. The descending curves represent counting-statistics MDC under idealized exact-baseline conditions; horizontal colored dash-dot lines indicate the systematic detection floor corresponding to an illustrative ±1% baseline-reference uncertainty (blue: ~28 ppm for aperture ROI; red: ~22 ppm for GTV ROI). Beyond the intersection of the descending curve and the horizontal floor, baseline uncertainty

dominates over counting statistics. Horizontal gray dotted line: 20 ppm reporting reference.

In practice, the baseline reference — whether derived from simulation, pre-treatment calibration, or phantom measurement — carries its own uncertainty. To illustrate the impact of baseline-reference precision, a ±1% uncertainty is adopted as a reference level. At this level, the resulting systematic detection floor — independent of integration time — is approximately 28 ppm for the aperture ROI (= 1%/0.036 %/ppm) and 22 ppm for the GTV ROI (= 1%/0.046 %/ppm). This floor dominates the MDC at integration times beyond approximately 2–3 s (Fig. 12), where the counting-statistics curve falls below the baseline-limited floor. The 20 ppm reporting reference lies below this floor; detecting a 20 ppm concentration difference would require baseline-reference precision better than ±1%. The ±1% level is not derived from a specific detector technology but serves as a scaling reference: the floor scales linearly with baseline uncertainty (e.g., ±3% would yield approximately 66–83 ppm). At clinically relevant concentrations (≥60 ppm, aperture RDSR ≈ 3.8%), the baseline contribution has proportionally less impact.

The separation between counting-statistics MDC (4.8 ppm) and baseline-uncertainty floor (22–28 ppm) demonstrates that the backscattered signal carries sufficient boron-concentration information; the system-level detection limit is governed by baseline-reference precision, not by signal strength or detector counting statistics. This distinction informs the development priority: detector optimization should target count-rate improvement and enrichment, while the baseline-acquisition strategy (Section 3.7) represents the dominant challenge for achieving clinically relevant detection thresholds.

Electronics-level parameters not modeled in this study — gas gain, threshold, charge sharing, dead time, and dark counts — will affect the achievable count rate and therefore the counting-statistics MDC. However, because RDSR is a baseline-normalized ratio, detector-side factors that act as spatially uniform multiplicative gains — gas gain, electronic efficiency, $^{6}$Li areal density — cancel to first order in the thin-converter, linear-response, spatially-uniform-gain regime considered here, i.e., when pile-up, dead-time saturation, and gain nonuniformity are negligible. Under these conditions, changing such parameters scales the absolute count rate rather than the fractional boron-dependent contrast. This separates the contrast problem, governed primarily by patient-side transport and $^{10}$B absorption, from the sensitivity problem, governed by count rate and integration time. Detector optimization should therefore prioritize count-rate improvement while preserving the thermal neutron energy weighting that produces the RDSR-to-RFR enhancement.

### *3.6 Neutron transparency and dosimetric impact of the modeled detector*

Fig. 13(a) compares DVHs for all scored regions — GTV, brain, skin, and other tissues — between the reference plan without the detector (solid lines) and the plan with the detector inserted at the collimator exit (dotted lines). Both plans were normalized to the same prescription constraint (maximum normal-brain dose of 12 Gy-Eq at 20 mA). After normalization, the curves for each ROI are nearly indistinguishable: the GTV distribution remains concentrated in the 55–65 Gy-Eq range, brain dose stays predominantly below 12 Gy-Eq, and skin and other tissue curves retain their characteristic low-dose profiles.

The detector produced limited DVH perturbation after prescription normalization, at the cost of a treatment-time increase from 54.30 min (without detector) to 60.60 min (with detector), an increment of +11.6%. This increase reflects beam attenuation by the detector materials and represents the primary dosimetric trade-off of the proposed measurement geometry. The small dosimetric perturbation reflects the neutron transparency of the modeled detector: the thin $^{nat}$LiF conversion layer (15 nm), low-Z structural materials, and millimeter-scale total thickness produce limited attenuation of the therapeutic neutron beam. Any detector implementation for this measurement concept would be subject to similar neutron-transparency constraint. In current clinical practice, irradiation time is routinely individualized — determined by patient-specific blood boron concentration measured immediately before treatment and adjusted for the prescription dose constraint of each organ at risk Takeno et al. (2024). The +11.6% treatment-time increase from the modeled detector (6.3 min at 20 mA) falls within the range of routine per-patient adjustments in this individualized treatment paradigm. After prescription-dose renormalization, ROI-level DVH metrics showed limited perturbation (GTV $D_{\mathrm{mean}}$ +0.20%, $D_{95}$ −0.02%; Brain $D_{1\mathrm{cc}}$

+0.13%; Skin $D_{0.1\%}$ +2.76%). Whether the treatment-time trade-off is clinically acceptable requires formal clinical workflow assessment.

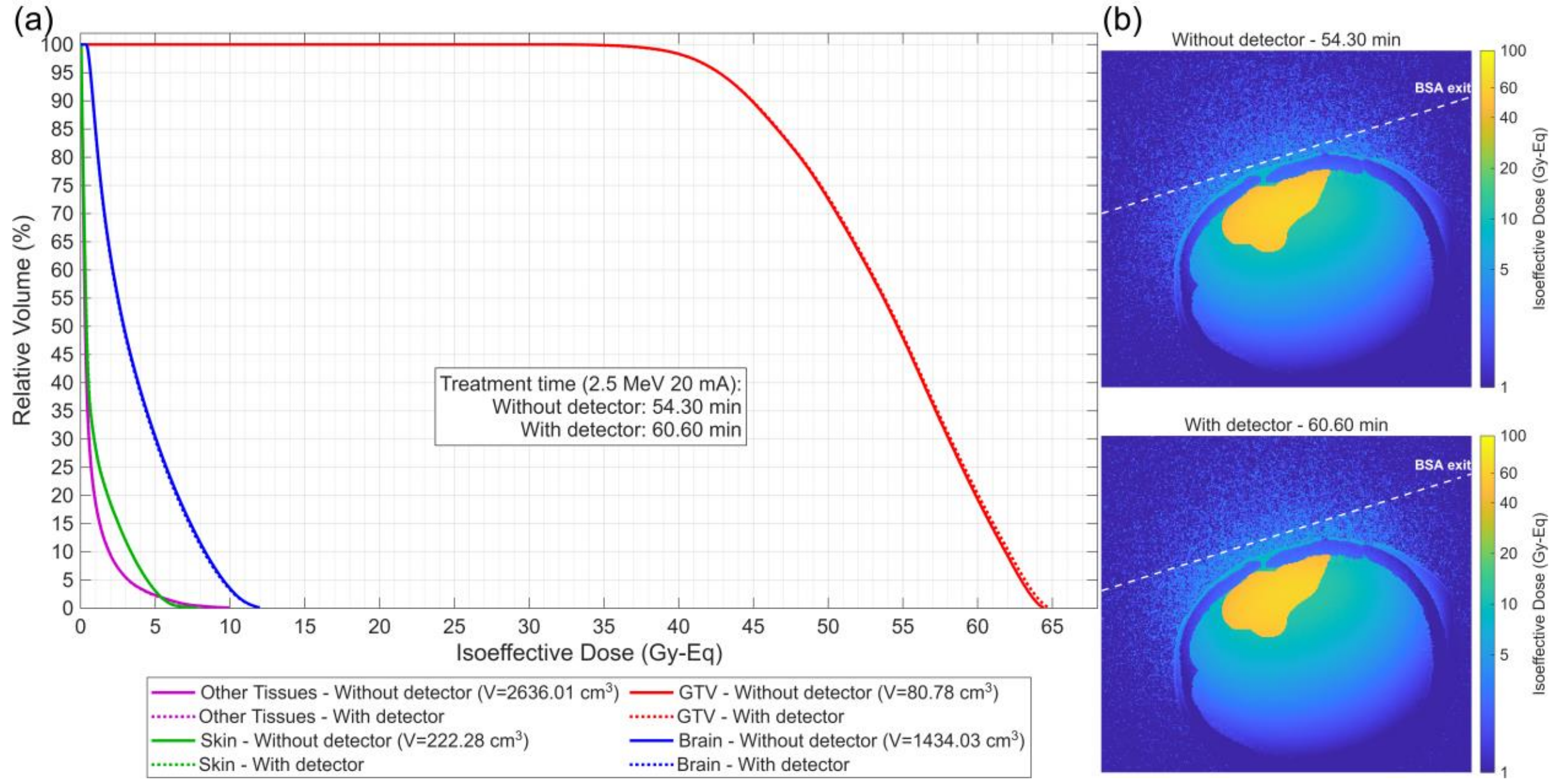

**Fig. 13.** Dosimetric impact of the modeled detector. (a) DVH comparison for GTV (red, $V = 80.78\ \text{cm}^3$), brain (blue, $V = 1434.03\ \text{cm}^3$), skin (green, $V = 222.28\ \text{cm}^3$), and other tissues (magenta, $V = 2636.01\ \text{cm}^3$). Solid: without detector; dotted: with detector. Both normalized to maximum normal-brain dose of 12 Gy-Eq at 2.5 MeV, 20 mA. Treatment time: 54.30 min (without) and 60.60 min (with detector). (b) Coronal isoeffective dose distributions (logarithmic scale, 1–100 Gy-Eq).

### *3.7 Limitations*

This study has several limitations.

*Single-patient scope.* All patient-specific findings derive from one voxelized head phantom. The 4.9-fold sensitivity enhancement over the homogeneous phantom (Section 3.2) is specific to this anatomy and cannot be assumed transferable. Multi-patient generalization studies are necessary to map the range of expected RDSR sensitivity across diverse anatomies.

*Simulation-only framework.* No experimental prototype has been constructed or tested. The validation benchmarks described in Section 2.2 support the physical trends but do not validate the end-to-end measurement chain. Source spectral and angular distribution variations would affect the forward/backscatter ratio and absolute RDSR and FWHM values, though relative concentration trends should be preserved.

*Baseline-reference strategy.* All RDSR values are computed against a Monte Carlo $^{10}$B-free baseline. In practice, the baseline must come from TPS-predicted simulation (CT-to-material conversion and source-model accuracy), exposure prior to boronophenylalanine (BPA) injection (positioning reproducibility and additional patient dose), or phantom calibration (anatomical mismatch). Any geometric change between baseline and treatment acquisitions — including patient positioning shifts — produces a spatially structured differential signal that may be indistinguishable from a concentration-dependent RDSR change. Positioning errors have notable dosimetric consequences in AB-BNCT: beam-axis displacement can change tumor $D_{80\%}$ by about 2.1 Gy-Eq/cm Kakino et al. (2022). As quantified in Section 3.5, achieving the ±1% baseline precision required for 22–28 ppm detection floors remains undemonstrated; the baseline-acquisition strategy is the dominant engineering challenge.

*Electronics and gamma background.* Gas gain, threshold, charge sharing, dead time, dark counts, and gamma discrimination efficiency were not modeled. These parameters affect absolute count yield and SNR but not the energy-selective weighting mechanism. The sub-second detectability estimates represent modeled performance bounds, not achievable specifications.

*Spatial resolution.* The edge-response characterization (Section 3.4) was performed in a homogeneous phantom; heterogeneous patient anatomy would modify the effective point-spread function. Patient-specific characterization is deferred to future work.

*Uncertainty budget.* The principal uncertainty sources are organized in Table 3. The GTV-projection ROI MDC assumes prior knowledge of tumor geometry from treatment-planning CT segmentation; the aperture ROI provides a prior-independent readout. A combined uncertainty is not reported; the RDSR values and detection benchmarks should be interpreted as modeled performance bounds.

**Table 3.** Principal uncertainty sources affecting the reported RDSR values and spatial-resolution estimates.

| # | Source | Effect on reported metrics | Status |
|---|---|---|---|
| 1 | MC statistical convergence | RSE < 0.05% (dose), < 0.02% (fluence) | Negligible |
| 2 | Nuclear data and physics model | Affects absolute RDSR; relative concentration trends preserved. Albedo benchmark within 1.1% of diffusion theory (Section 2.2) | Bounded by benchmark |
| 3 | Patient anatomy | Single voxelized phantom; sensitivity enhancement is anatomy-dependent | Characterized for one geometry |
| 4 | Boron distribution model | Uniform intra-region assumption; clinical heterogeneity modifies local RDSR | Parametrically varied (Table 2) |
| 5 | Detector response model | Electronics-level parameters not modeled; affect count yield, not energy-selective weighting | RDSR contrast preserved |
| 6 | Spatial resolution characterization | Error-function fit to finite-thickness slab in homogeneous medium; Gaussian LSF assumption | Upper-bound FWHM estimate |

## 4. Conclusions

We hypothesized that the backscattered thermal neutron field carries sufficient $^{10}$B-dependent information to serve as a viable measurement signal for boron concentration during the delivery of BNCT. To our knowledge, this is the first proof-of-concept assessment of this signal carrier — a physical phenomenon documented in BNCT treatment rooms for three decades but previously treated as a nuisance, correction term, or confounding factor. Geant4-based Monte Carlo simulations support this hypothesis across all four aspects investigated: (i) *concentration and depth sensitivity* — the detector-level signal (RDSR) exhibits linear concentration dependence with a practical depth limit of approximately 6 cm; (ii) *spatial resolution* — edge-response analysis yields a diffusion-limited FWHM of 32–176 mm over the 1–5 cm depth range, indicating centimeter-scale resolution governed by thermal neutron diffusion; (iii) *clinical-geometry sensitivity* — in the voxelized patient phantom, the concentration-response sensitivity substantially exceeded that of the homogeneous phantom, and ROI-integrated analysis indicated detectability at clinically relevant timescales under idealized baseline conditions; and (iv) *dosimetric perturbation* — the modeled detector increases treatment time by 11.6% with limited DVH perturbation after prescription normalization, consistent with the neutron-transparency design of the thin $^{nat}$LiF-converter detector model. The $^{6}$Li capture cross-section further provides intrinsic energy selectivity that preferentially weights the thermal band where $^{10}$B absorption is concentrated, without requiring external energy discrimination.

ROI-integrated analysis reveals that the limiting factor for detection sensitivity is baseline-reference precision rather than intrinsic signal strength, identifying the baseline-acquisition strategy as the dominant development challenge. These results provide quantitative design targets for three development priorities: (1) experimental prototype validation under accelerator beam conditions; (2) baseline-reference strategy development to achieve the sub-percent precision required for clinically relevant detection thresholds; and (3) multi-patient generalization across diverse anatomies.

The physical basis established here — that a thin, neutron-transparent detector at the beam exit is predicted to extract spatially resolved $^{10}$B-dependent information from the backscattered field — provides the foundation for a new radiation-measurement approach to real-time boron monitoring in BNCT.

## Acknowledgements

This work was supported in part by Anhui Province Key Research and Development Plan (2023s07020020), National Natural Science Foundation of China (12275372), the Fourth Batch of R&D Projects on Nuclear Technology under the 14th Five-Year Plan (HJSYF2025(22)), and Research Funds of the Joint Research Center for Regional Diseases of IHM (2023bydjk002).

## CRediT author contributions

**Zirui Ye:** Writing – original draft, Visualization, Validation, Software, Methodology, Investigation, Formal analysis, Data curation, Conceptualization. **Yuxin Wang:** Writing – review & editing, Software, Validation, Investigation, Data curation. **Meitong Wei:** Writing – review & editing, Investigation. **Xie George Xu:** Writing – review & editing, Supervision,

Methodology, Investigation, Resources, Project administration, Funding acquisition, Conceptualization.

## Declaration of competing interests

The authors declare no conflicts of interest.

## Data availability

Data are available from the corresponding author on reasonable request.